\documentstyle[12pt,epsf]{article}

\advance \topmargin by -\headheight
\advance \topmargin by -\headsep

\evensidemargin \oddsidemargin
\def\im{{\hbox{\rm Im}}}

\def\ie{{\em i.e.}}

\def\ie{\hbox{\it i.e.}}

\def\CC{{\mathchoice
{\rm C\mkern-8mu\vrule height1.45ex depth-.05ex 
width.05em\mkern9mu\kern-.05em}
{\rm C\mkern-8mu\vrule height1.45ex depth-.05ex 
width.05em\mkern9mu\kern-.05em}
{\rm C\mkern-8mu\vrule height1ex depth-.07ex 
width.035em\mkern9mu\kern-.035em}
{\rm C\mkern-8mu\vrule height.65ex depth-.1ex 
width.025em\mkern8mu\kern-.025em}}}

\def\RR{{\rm I\kern-1.6pt {\rm R}}}

\def\ZZ{{\rm Z}\kern-3.8pt {\rm Z} \kern2pt}

\def\np{Nucl. Phys.}
\def\pl{Phys. Lett.}

\def\cmp{Commun. Math. Phys.}
\def\jmp{J. Math. Phys.}
\def\ijmp{Int. J. Mod. Phys.}
\def\mpl{Mod. Phys. Lett.}
\def\lmp{Lett. Math. Phys.}

\def\jpsc{J. Phys. Soc. Jap.}
\def\topo{Topology}

\def\phyrep{Phys. Rep.}

\def\rmp{Rev. Math. Phys.}
\def\im{Invent. Math.}

\newcommand{\beq}{\begin{equation}}
\newcommand{\eeq}{\end{equation}}
\newcommand{\rc}{\nonumber\\}
\newcommand{\bear}{\begin{eqnarray}}
\newcommand{\eear}{\end{eqnarray}}

\newfont{\namefont}{cmr10}
\newfont{\addfont}{cmti7 scaled 1440}
\newfont{\boldmathfont}{cmbx10}
\newfont{\headfontb}{cmbx10 scaled 1728}
\renewcommand{\theequation}{{\rm\thesection.\arabic{equation}}}
\begin{document}
\begin{titlepage}

\begin{center} \Large \bf Duality in 
osp$\bf{(1\vert 2)}$ Conformal Field Theory 
\\ and link invariants

\end{center}

\vskip 0.3truein
\begin{center} 
I.P. Ennes${}^{\,\dagger}$
\footnote{e-mail:ennes@gaes.usc.es}, 
P. Ramadevi${}^{\,*}$
\footnote{e-mail:rama@theory.tifr.res.in},  
A.V. Ramallo${}^{\,\dagger}$
\footnote{e-mail:alfonso@gaes.usc.es}
and 
J. M. Sanchez de Santos ${}^{\,\dagger}$
\footnote{e-mail:santos@gaes.usc.es}

\vspace{0.3in}

${}^{\dagger\,}$Departamento de F\'\i sica de
Part\'\i culas, \\ Universidad de Santiago\\
E-15706 Santiago de Compostela, Spain. 
\vspace{0.3in}

${}^{\,*}$Tata Institute of Fundamental Research, \\
Homi Bhabha Road, Mumbai, India 400005.

\end{center}
\vskip 1truein

\begin{center}
\bf ABSTRACT
\end{center} 

We study the crossing symmetry of the conformal blocks of
the conformal field theory based on the affine Lie
superalgebra osp$(1\vert 2)$. Within the framework of a free
field realization of the osp$(1\vert 2)$ current algebra,
the fusion and braiding matrices of the model are
determined. These results are related in a simple way to
those corresponding to the su$(2)$ algebra by means of a
suitable identification of parameters. In order to obtain
the link invariants corresponding to the osp$(1\vert 2)$
conformal field theory, we analyze the corresponding
topological Chern-Simons theory. In a first approach we
quantize the Chern-Simons theory on the torus and, as a
result, we get the action of the Wilson line operators on
the supercharacters of the affine osp$(1\vert 2)$. From
this result we get a simple expression relating the 
osp$(1\vert 2)$ polynomials for torus knots and links to
those corresponding to the su$(2)$ algebra. Further, this
relation  is verified for arbitrary knots and links by 
quantizing the Chern-Simons theory on the punctured
two-sphere.

\vskip1.5truecm
\leftline{US-FT-27/97 \hfill August 1997}
\leftline{TIFR/TH/97-35}
\leftline{hep-th/9709068}
\smallskip
\end{titlepage}
\setcounter{footnote}{0}

\setcounter{equation}{0}
\section{Introduction}

The study of the duality properties of Conformal Field
Theory (CFT) has provided interesting information about the
global structure of these theories \cite{MS, Review}. 
Indeed, the behaviour
of the conformal blocks under crossing symmetry determines
a consistent realization of a non-trivial exchange algebra,
whose analysis has revealed the existence of a hidden
quantum group symmetry \cite{AGS}. This quantum group
symmetry can be used to define new invariants for knots and
links in three-dimensional topology \cite{KR,Res}. 

The topological information encoded in the duality structure
of CFT can be directly extracted by formulating a topological
Chern-Simons gauge theory in three dimensions
\cite{witten}.  The states in the Chern-Simons theory can be
identified with the conformal blocks of the two-dimensional
CFT. The basic observables in the Chern-Simons theory are
the Wilson lines, which are invariant under both gauge
symmetry and topological deformations. From their vacuum
expectation values, which can be obtained by exploiting the
correspondence with CFT, one can define topological
invariants for knots and links. In this way, a connection
between two-dimensional field theories and three-dimensional
topology can be neatly established. 

In this paper we shall carry out the analysis, along the
lines described above, of the CFT based on the affine Lie
superalgebra osp$(1\vert 2)$. This CFT shows up in
many problems in which the N=1 superconformal symmetry is
present \cite{ber,polyakov}.  
We have recently studied \cite{osp} the structure of its
conformal blocks by means of a free field realization. In
the present work, we shall use the representation of the
conformal blocks found in ref. \cite{osp} in order to
determine their behaviour under the different crossing
symmetry  transformations. Making use of the contour
manipulation techniques,  one can obtain the braiding and
fusion matrices from the free field representation of the
osp$(1\vert 2)$  CFT. We shall discover a great similarity
between the duality matrices of osp$(1\vert 2)$ and those
corresponding to the su$(2)$ CFT. Actually, we shall find
that, by performing a suitable identification, 
our osp$(1\vert 2)$ duality matrices can be obtained
from the su$(2)$ ones.

Armed with these results for the braiding and fusion
structure of the osp$(1\vert 2)$ CFT, one can try to find
out what kind of invariants for knots and links are
generated by this model. Knot invariants associated to Lie
superalgebras have been considered from different points of
view in refs. \cite{Horne,KS,Rozansky,Deguchi,Zhang}. 
Here we shall formulate a Chern-Simons
topological theory with  osp$(1\vert 2)$ as gauge symmetry.
In order to obtain the corresponding knot polynomials we
shall apply the methods of refs. \cite{LR,kaul} , 
which allow to perform
a direct evaluation of the invariants. The final result of
our analysis is an equation that relates the 
osp$(1\vert 2)$ and su$(2)$ polynomials. This relation is
of the same type as the one found between the duality
matrices. 

This paper is organized as follows. In section 2 we shall
recall some basic facts of the osp$(1\vert 2)$ CFT, in
particular its fusion algebra and the expression of its
supercharacters. We shall also study in this section the
modular transformations of the osp$(1\vert 2)$
supercharacters and we shall relate the values of the
modular $S$ matrix to the quantum dimensions appearing in
the quantum deformation of the osp$(1\vert 2)$ symmetry. In
the course of this analysis, we shall find an identification
between the deformation parameters of osp$(1\vert 2)$ and 
su$(2)$ which allows to write the osp$(1\vert 2)$ modular
$S$ matrix in terms of  su$(2)$ q-numbers. This 
osp$(1\vert 2)$/su$(2)$ identification is precisely the one
we shall find for the duality matrices and knot
polynomials. 

In section 3 we shall implement the crossing symmetry 
operations in the free field representation of the 
osp$(1\vert 2)$ CFT and, as a result, we shall be able to
find general expressions for the braiding and 
fusion matrices. Following the approach of ref. \cite{osp},
we shall express the conformal blocks as multiple contour
integrals. The monodromy properties of these integrals can
be studied with the techniques introduced in ref. \cite{DF}
and, as a result, the fusion and braiding matrices
of the  osp$(1\vert 2)$ CFT can be determined. 

In section 4 we develop the operator
formalism of ref. \cite{LR} for the osp$(1\vert 2)$
Chern-Simons gauge theory. In this formalism it is possible 
to represent the Wilson lines as operators  acting on the
supercharacters of the two-dimensional theory. Actually, one
has to deal with an effective quantum mechanical problem
whose corresponding Hilbert space is finite dimensional. The
states in this effective problem are the supercharacters and
the observables, \ie\ the Wilson lines, are represented as
differential operators acting on them. Using
this representation, the form of the Verlinde operators 
\cite{Verlinde} can be obtained. Moreover,  
the expectation values of Wilson lines
for torus knots and links can be easily calculated and the
expression of the corresponding invariant polynomials can
be determined. From these results for torus knots and links
we obtain the above-mentioned relation between the 
osp$(1\vert 2)$ and su$(2)$ invariant polynomials. 

In order
to extend this analysis to more general classes of knots
and links, we shall adopt in section 5 the approach of ref.
\cite{kaul}, in which the quantization surface is the
two-sphere with punctures. The basic input one needs in this
approach is  behaviour under braiding and fusion of the 
osp$(1\vert 2)$ CFT, which was obtained in section 3. 
Using this formalism one can obtain systematically the
invariant polynomials from some basic states of the
Chern-Simons Hilbert space on the punctured two-sphere. The
results  we shall find with these methods confirm the
relation between the  osp$(1\vert 2)$ and su$(2)$ 
polynomials found in section 4. Finally, in section 6 we
recapitulate our results and present some conclusions. Some
details of our calculations are contained in three
appendices.

\setcounter{equation}{0}
\section{  osp$\bf{(1\vert 2)}$ Conformal Field Theory }

The affine ${\rm osp}(1\vert 2)$ Lie superalgebra is 
generated by three bosonic currents, which we shall denote
by $J^{\pm}(z)$ and $J^{0}(z)$, and by two fermionic
operators $j^{\pm}(z)$. These operators can be expanded in
modes as follows:
\beq
J^{a}(z)\,=\,\sum_{n\in\ZZ}\,\,
J^{a}_n\,z^{-n-1}
\,\,\,\,\,\,\,\,\,\,\,\,\,\,\,\,\,\,\,\,
j^{\alpha}(z)\,=\,\sum_{n\in\ZZ}\,\,
j^{\alpha}_n\,z^{-n-1}\,\,.
\label{uno}
\eeq
The affine ${\rm osp}(1\vert 2)$ is defined by the set of
(anti)commutators \cite{Review}:
\bear
&&[\,J_n^0\,,\,J_m^{\pm}\,]\,=\,\pm J_{n+m}^{\pm}
\,\,\,\,\,\,\,\,\,\,\,\,\,\,\,\,\,\,\,\,\,\,\,\,\,
\,\,\,\,\,\,\,\,\,\,\,\,\,\,\,\,\,\,\,\,\,\,\,\,\,
[\,J_n^0\,,\,J_m^{0}\,]\,=\,
{k\over 2}\,n\,\delta_{n+m}\rc
&&[\,J_n^+\,,\,J_m^{-}\,]\,=\,kn\delta_{n+m}\,+\,
2J^0_{n+m}\rc
&&[\,J_n^0\,,\,j_m^{\pm}\,]\,=\,\pm\,{1\over 2}\,
j_{m+n}^{\pm}
\,\,\,\,\,\,\,\,\,\,\,\,\,\,\,\,\,\,\,\,\,\,\,\,\,
\,\,\,\,\,\,\,\,\,\,\,\,\,\,\,\,\,\,\,\,
[\,J_n^{\pm}\,,\,j_m^{\pm}\,]\,=\,0\label{dos}\\
&&[\,J_n^{\pm}\,,\,j_m^{\mp}\,]\,=\,-j_{n+m}^{\pm}
\,\,\,\,\,\,\,\,\,\,\,\,\,\,\,\,\,\,\,\,\,\,\,\,\,
\,\,\,\,\,\,\,\,\,\,\,\,\,\,\,\,\,\,\,\,\,\,\,\,\,
\{\,j_n^{\pm}\,,\,j_m^{\pm}\,\}\,=
\,\pm 2 J_{n+m}^{\pm}\rc
&&\{\,j_n^+\,,\,j_m^{-}\,\}\,=\,2kn\delta_{n+m}\,+\,
2J^0_{n+m}\,\,.\rc
\nonumber
\eear

In eq. (\ref{dos}),  $k$ is a c-number (the level of the
algebra), which we will take to be a non-negative integer.
Based on the algebra (\ref{dos}), one can construct a CFT
whose energy-momentum tensor is given by the Sugawara
expression:
\bear
T(z)\,=\,{1\over 2k+3}\,&:\,[\,2\,(J^0(z))^2\,+\,
J^+(z)\,J^-(z)\,+\,
J^-(z)\,J^+(z)\,\nonumber\\
&-\,{1\over 2}\,j^+(z)\,j^-(z)\,+\,
{1\over 2}\,j^-(z)\,j^+(z)\,]:\,\,.
\label{tres}\\
\nonumber
\eear
The central charge corresponding to the operator $T$ in
(\ref{tres}) is:
\beq
c\,=\,{2k\over 2k+3}\,\,.
\label{cuatro}
\eeq

The zero-mode operators $J_{0}^{a}$ and $j_{0}^{\alpha}$
satisfy a finite dimensional (\ie\ non-affine) 
${\rm osp}(1\vert 2)$ superalgebra (see eq. (\ref{dos})). The
finite-dimensional representations of this superalgebra are
characterized by an integer or half-integer number $j$,
which we shall refer to as the isospin of the
representation. The state of the representation with
eigenvalue $m$ with respect to the Cartan generator
$J_{0}^{0}$ will be denoted by $|j, m>$, being $|j, j>$ the
highest weight state.  Acting with the odd operators
$j_{0}^{\pm}$, the  $J_{0}^{0}$-eigenvalue of the state is
shifted by $\pm 1/2$ and, as a consequence, $m$ can take the
values  $j, j-{1\over 2}, \cdots, -j+{1\over 2}, -j$. Thus,
the isospin $j$ representation is $4j+1$-dimensional. The
statistics of the different states of the representation is
determined by the Grassmann parity $\lambda$ of the highest
weight state  $|j, j>$.   When $|j, j>$ is bosonic
(fermionic) the parameter $\lambda$, which we simply call
the parity of the representation, takes the value 0(1) and
we will say that the representation  is even (odd). It is
clear that the state $|j, m>$ is bosonic(fermionic) if 
$\lambda\,+\,2(j-m)$ is zero(one) modulo two. 

The  similarity between  the ${\rm osp}(1\vert 2)$ and
su$(2)$ representation theory is quite evident. Let us
point out, however, an important difference. In 
an ${\rm osp}(1\vert 2)$  representation, one can have states
with negative norm and, therefore, the corresponding CFT is
not unitary. This fact is related to the generalization of
the adjoint operation that one must adopt for this graded
algebra \cite{Pais}.  We shall choose our conventions in
such a way that the norm of the state $|j, m>$ is 
$(-1)^{2\lambda(j-m)}$. Therefore, only those states
belonging to an odd representation (\ie\ with $\lambda=1$)
and with $j-m\in\ZZ+{1\over2}$, will have negative norm.

The primary fields of the ${\rm osp}(1\vert 2)$ CFT are
associated to the states $|j, m>$ of the finite algebra
described above. Let us denote by $\Phi_m^j$ the primary
field corresponding to the state $|j, m>$. The conformal
weight of one of such operator in terms of  the quadratic
Casimir $c_j$ of the representation is given by:
\beq
h_j\,=\,{2c_j\over 2k+3}\,=\,
{j\,(\,2j\,+\,1\,)\over 2k\,+\,3}\,,
\label{cinco}
\eeq
where $c_j=j(j+{1\over 2})$. Notice that, as it should, the
conformal weights (\ref{cinco}) of the fields $\Phi_m^j$ do
not depend on $J_{0}^{0}$-eigenvalue $m$ and, thus, we
have a $4j+1$-dimensional multiplet of fields associated to
each value of the isospin. The algebra satisfied by these
operators is encoded in the fusion rules of the model,
which can be obtained from the selection rules of its 
operator algebra \cite{osp,yudos}. If $[j;\lambda]$
denotes the representation with isospin $j$ and parity
$\lambda$, the fusion rules read:
\beq
[j_1;\lambda_1]\,\times\,[j_2;\lambda_2]\,=\,
\sum_{{j_3=|j_1-j_2|\atop}\atop 2(j_3-j_1-j_2)\,\in\,\ZZ}
^{{\rm min}\,(\,j_1+j_2\,,\,k+{1\over 2}-j_1-j_2\,)}
\,\,\,[j_3;\lambda_3]\,\,,
\label{seis}
\eeq
where $\lambda_3$ is related to $\lambda_1$ and $\lambda_2$
as follows:

\beq
\lambda_3\,=\,\lambda_1\,+\,\lambda_2\,+\,2(j_1+j_2-j_3)
\,\,\,\,\,\,\,\,\,\,\,\,\,\,\,\,\,
{\rm mod}\,(2)\,\,.
\label{siete}
\eeq

Again, the similarity with su$(2)$ is manifest in the
composition law (\ref{seis}). Indeed, it follows from 
(\ref{seis}) that the space of fields with $j\le k/2$ is
closed under multiplication. However, the coupling of two
isospins $j_1$ and $j_2$ gives rise to isospins $j_3$ such
that $j_3-j_1-j_2$ is integer or half integer. Notice that, 
in this last case, the parity $\lambda_3$ is not the sum of
$\lambda_1$ and $\lambda_2$ (see eq. (\ref{siete})) and, as
a consequence, one can get odd representations by coupling
two even ones.

The supercharacters for the spin $j$ representation 
of the  ${\rm osp}(1\vert 2)$ current algebra at
level $k$ are defined as:

\beq
\chi_{j,k}(a,\tau)\,=\,{\rm Str}_j\,
[\,e^{2\pi i \tau(L_0-{c\over 24})}\,
e^{2\pi i aJ_0^0}\,]\,,
\label{ocho}
\eeq
where $\tau $ is the modular parameter, $L_0$ is
the zero mode part of the energy-momentum tensor $T$ and
$a$ is a variable associated to the Cartan generator
$J_0^0$. In eq. (\ref{ocho}),  ${\rm Str}_j$ denotes the
supertrace over the Verma module whose highest weight state
is $|j, j>$, \ie\ the trace over the bosonic states minus
the trace over the fermionic states of the Verma module. 
The explicit form of the functions 
$\chi_{j,k}(a,\tau)$ has been obtained in ref.
\cite{yudos}.  They can be written in terms of the functions:

\beq
\vartheta_{j,k}(a,\tau)\,\equiv\,\Theta_{j,k}\,
(\,{a+1\over 2}\,,{\tau\over 2}\,)\,\,,
\label{nueve}
\eeq
where $\Theta_{j,k}(a, \tau)$ are classical theta
functions. From the definition (\ref{nueve}) one can easily
conclude that the functions $\vartheta_{j,k}(a,\tau)$ can be
represented by the series:

\beq
\vartheta_{j,k}(a,\tau)\,=\,e^{{i\pi\over 2}j}\,
\sum_{n\in\ZZ}\,e^{i\pi k\tau(n+{j\over 2k})^2}\,
e^{i\pi ka(n+{j\over 2k})+i\pi kn}\,\,\,.
\label{diez}
\eeq
It follows from (\ref{diez}) that $\vartheta_{j,k}(a,\tau)$
satisfies $\vartheta_{j,k}(a,\tau)\,=\,\vartheta_{j\pm
2k,k}(a,\tau)$. For an even representation, the 
$\chi_{j,k}(a,\tau)$ character is given in terms of 
$\vartheta_{j,k}(a,\tau)$ by means of the expression
\cite{yudos}:

\beq
\chi_{j,k}(a,\tau)\,=\,
{\vartheta_{4j+1,2k+3}(a,\tau)\,-\,
\vartheta_{-4j-1,2k+3}(a,\tau)\over
\Pi(a,\tau)}\,\,,
\label{once}
\eeq
with:
\beq
\Pi(a,\tau)\,=\,\vartheta_{1,3}(a,\tau)\,-\,
\vartheta_{-1,3}(a,\tau)\,\,.
\label{doce}
\eeq

From the periodicity properties of the $\vartheta_{j,k}$
functions, it follows that the supercharacters $\chi_{j,k}$
satisfy:

\bear
\chi_{j,k}(a,\tau)&=&
-\chi_{-j-{1\over 2},k}(a,\tau)\rc
\chi_{k+1-j,k}(a,\tau)&=&
-\chi_{j,k}(a,\tau)\,\,.
\label{trece}
\eear
It is immediate from (\ref{trece}) that there are only $k+1$
 independent supercharacters $\chi_{j,k}$, which correspond
to the isospins 
$j=0, {1\over 2},\cdots,{k\over 2}$, in agreement with the
fusion rule (\ref{seis}). Performing a Poisson ressumation,
one can get the behaviour of the supercharacters under 
modular transformations $\tau\rightarrow
-{1\over\tau}$ and $a \rightarrow {a\over \tau}$. The
result is:

\beq
\chi_{j,k}({a\over \tau},-{1\over\tau})\,=\,
e^{{i\pi k\over 2\tau}\,a^2}\,\,
\sum_{{l=0\atop}\atop 2l\,\in\,\ZZ}^{k/2}
S_{jl}\,\,\chi_{l,k}(a,\tau)\,\,,
\label{catorce}
\eeq
where the $S$ matrix is given by:
\beq
S_{jl}\,=\,\sqrt{4\over 2k+3}\,\,(-1)^{2j+2l}\,
{\rm cos}\,\Bigl[\,{(4j+1)(4l+1)\over
2(2k+3)}\pi\,\Bigr]\,\,.
\label{quince}
\eeq

It is clear from (\ref{catorce}) that the $S$-matrix 
(\ref{quince}) determines the transformation of the
specialized  supercharacters $\chi_{l,k}(0,\tau)$ under
the   $\tau\rightarrow-{1\over\tau}$  transformation. 
\footnote{Alternatively, the exponential factor in the
right-hand side of (\ref{catorce}) can be absorbed by
multiplying the supercharacters by a convenient prefactor} 
It is important to point out the appearance in
(\ref{quince}) of a cosine, instead of the usual sine that
one has in the su$(2)$ theory. It is also interesting to 
analyze the $S$-matrix ratios $S_{0j}/S_{00}$. Indeed,
according to general arguments in CFT, these ratios
should be related to the quantum dimensions \cite{DV} of some
representations of  the q-deformation of the universal
enveloping algebra of ${\rm osp}(1\vert 2)$. Actually
\cite{DV}, these quantities determine the ratio
$\chi_{j,k}/\chi_{0,k}$ between the isospin $j$
supercharacter and that of the vacuum in the
$\tau\rightarrow 0$ limit. In order to give an
interpretation of these $S$-matrix quotients in our case,
let us introduce the quantity:
\beq
q\,=\,{\rm exp}\,[\,{i\pi\over 2k+3}\,]\,\,.
\label{dseis}
\eeq
On the other hand, the graded ${\rm osp}(1\vert 2)$ 
q-numbers, denoted by $[x]_+$ for $x$ integer or
half-integer,  are defined as \cite{Kulis,Saleur}:
\beq
[x]_+\,=\,{q^{x}\,-\,(-1)^{2x}\,q^{-x}\over
q^{{1\over 2}}\,+\,q^{-{1\over 2}}}\,\,.
\label{dsiete}
\eeq
Using the definitions (\ref{dseis}) and (\ref{dsiete}), it
is straightforward to write $S_{0j}/S_{00}$ as:
\beq
{S_{0j}\over S_{00}}\,=\,(-1)^{2j}\,
\Bigl[\,{4j+1\over2}\,\Bigr]_+\,\,.
\label{docho}
\eeq
From eq. (\ref{docho}), one can easily verify that the
quotients $S_{0j}/S_{00}$ satisfy the fusion rules
(\ref{seis}) for the isospins, which is nothing but the
Verlinde theorem for our ${\rm osp}(1\vert 2)$ CFT.

Let us now see how the values (\ref{docho}) are related to
the quantum deformation of the ${\rm osp}(1\vert 2)$
symmetry. The universal enveloping algebra of 
${\rm osp}(1\vert 2)$ is generated by two fermionic
operators $F^{\pm}$ (corresponding to our currents 
$j^{\pm}$) and by a Cartan element $H$ (related to our
bosonic current $J^0$). Its q-deformation, which we shall
denote by $U_q\,({\rm osp}(1\vert 2))$, was introduced in
refs. \cite{Kulis,Saleur}. Its defining (anti)commutators
are:
\bear
[\,H\,,\,F^{\pm}\,]\,&=&\,\pm\,{1\over 2}\,\,F^{\pm}\rc\rc
\{\,F^{+}\,,\,F^{-}\,\}\,&=&\,
{q^{2H}\,-\,q^{-2H}\over q\,-\,q^{-1}}\,\,.
\label{dnueve}\\
\nonumber
\eear
There is a one-to-one relation between 
the irreducible representations
of (\ref{dnueve}) and those of the ${\rm osp}(1\vert 2)$
finite algebra. Actually, for every integer or half-integer
isospin $j$, there exists a $4j+1$-dimensional
representation in which the $H$ generator is diagonal with
eigenvalues $j, j-{1\over 2}, \cdots, -j+{1\over 2}, -j$.
As in the undeformed algebra, the representations are
referred to as even or odd  depending
on the parity of their highest weight vector. The super
q-dimension for a representation of isospin $j$ is defined
as:
\beq
SD_q[j]\,\equiv\,{\rm Str}_j\,q^{2H}\,\,,
\label{veinte}
\eeq
where now, ${\rm Str}_j$ denotes the supertrace over the
isospin $j$ representation of 
$U_q\,({\rm osp}(1\vert 2))$. It is a simple exercise to
compute the value of
$SD_q[j]$ for an even representation. The result is:
\beq
SD_q[j]\,=\,
\sum_{m=-j\atop 2m\in\,\ZZ}^{j}\,
(-1)^{2(j-m)}\,q^{2m}\,=\,
\Bigl[\,{4j+1\over2}\,\Bigr]_+\,\,.
\label{vuno}
\eeq
Therefore, we can rewrite (\ref{docho}) as:
\beq
{S_{0j}\over S_{00}}\,=\,(-1)^{2j}\,SD_q[j]\,\,.
\label{vdos}
\eeq

To finish this section, let us point out that there exist a
relation between the q-numbers (\ref{dsiete}) and those of
su$(2)$. If the deformation parameter of su$(2)$ is denoted
by $t$, we shall define $[x]$ as:
\beq
[x]\,=\,{t^{{x\over 2}}\,-\,t^{-{x\over 2}}\over
t^{{1\over 2}}\,-\,t^{-{1\over 2}}}\,\,.
\label{vtres}
\eeq
Suppose that $x\in\ZZ$ and that we identify $t=-q$. As
$(-1)^{{x\over 2}}\,=\,(-1)^x\,\,(-1)^{-{x\over 2}}$ for
any $x\,\in\,\ZZ$, one can write:
\beq
[x]\,=\,(-1)^{{x-1\over 2}}\,\,
\Bigl[\,{x\over 2}\,\Bigr]_+
\,\,\,\,\,\,\,\,\,\,\,\,\,\,\,
{\rm for}\,\,\,x\in\ZZ\,\,\,{\rm and}\,\,\,t=-q\,\,.
\label{vcuatro}
\eeq
Therefore we can rewrite eq. (\ref{docho}) as:
\beq
{S_{0j}\over S_{00}}\,=\,[4j+1].
\label{vcinco}
\eeq
Notice that the right-hand side of eq. (\ref{vcinco}) is
the quantum dimension corresponding to a representation of 
$U_t({\rm su}(2))$ with isospin $2j$. This relation between
the quantum groups based on su$(2)$ and 
${\rm osp}(1\vert 2)$, when one properly identifies their
deformation parameters, has been pointed out previously in
ref. \cite{Saleur} and will play an important r\^ ole in our
approach.

\setcounter{equation}{0}
\section{Braiding and Fusion in  osp$\bf{(1\vert 2)}$ CFT }

In this section we shall study the behaviour of the
conformal blocks of the osp$(1\vert 2)$ CFT under the
different exchanges of fields and duality transformations.
We shall restrict  our analysis to the case of the four
point conformal blocks, which will generally be denoted by 
${\cal F}^{1234}$. In the s-channel basis, the blocks can
be represented as in figure 1a, where $p$ is an
index labelling the s-channel intermediate state. The
s-channel block depicted in figure 1a will be denoted by 
${}^s{\cal F}^{1234}_p$, where the labels $1,2,3$ and $4$
represent both the quantum numbers of the primary fields
involved in the four point correlator and their locations.
There are different exchange operations that one can
perform on ${}^s{\cal F}^{1234}_p$. The simplest ones are
the interchange of the legs $1$ and $2$ ($3$ and $4$) of
the block, which will be denoted by $\pi_{12}$($\pi_{34}$):
\beq
\pi_{12}\,\Bigl[\,{}^s{\cal F}^{1234}_p\,\Bigr]\,=\,
{}^s{\cal F}^{2134}_p
\,\,\,\,\,\,\,\,\,\,\,\,\,\,\,\,\,\,\,\,\,\,\,\,\,\,
\,\,\,\,\,\,\,\,\,
\pi_{34}\,\Bigl[\,{}^s{\cal F}^{1234}_p\,\Bigr]\,=\,
{}^s{\cal F}^{1243}_p\,\,.
\label{vseis}
\eeq
\begin{figure}
\centerline{\hskip -.8in \epsffile{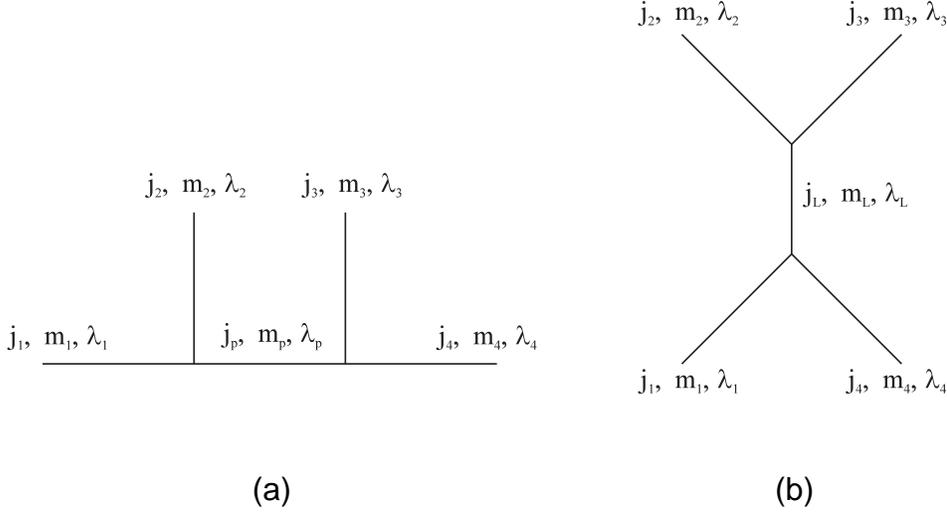}}
\caption{Diagrammatic representation of 
the conformal blocks for the s-channel (a) and for the
t-channel (b).}
\label{fig1}
\end{figure}
Notice (see figure 1a) that the fields $1$ and $2$ are
attached to the same three-point vertex in 
${}^s{\cal F}^{1234}_p$. Therefore one can assure that 
$\pi_{12}$ acts diagonally on the s-channel blocks. The same
conclusion is valid for $\pi_{34}$ and, thus, we can write:
\beq
\pi_{12}\,\Bigl[\,{}^s{\cal F}^{1234}_p\,\Bigr]\,=\,
\Lambda^{p}_{1,2}
\,\,\,{}^s{\cal F}^{1234}_p
\,\,\,\,\,\,\,\,\,\,\,\,\,\,\,\,\,\,\,\,\,\,\,\,\,\,
\,\,\,\,\,\,\,\,\,\,\,\,\,\,\,\,\,\,\,\,\,\,\,\,\,\,
\pi_{34}\,\Bigl[\,{}^s{\cal F}^{1234}_p\,\Bigr]\,=\,
\Lambda^{p}_{3,4}
\,\,\,{}^s{\cal F}^{1234}_p\,\,,
\label{vsiete}
\eeq
where $\Lambda^{p}_{1,2}$ and $\Lambda^{p}_{3,4}$ are
constants. We can also exchange the fields located at
positions $2$ and $3$. This is equivalent to the crossing 
symmetry between the s and u channels. Based on this s-u
duality, one can write:
\beq
{}^s{\cal F}^{1234}_p\,\,=\,\,
\sum_{l}\,\,B_{j_p,j_l}\Bigl[\,{2\,\,3\atop
1\,\,4}\Bigr]\,\,  {}^s{\cal F}^{1324}_l\,\,,
\label{vocho}
\eeq
where  
$B_{j_p,j_l}\Bigl[\,{2\,\,3\atop 1\,\,4}\Bigr]$ are the
elements of the so-called 
braiding matrix \cite{MS,Review}. Notice that the
different elements of this matrix are labelled by the
isospins of the intermediate channels. 

One can also use a t-channel basis to describe the space of
four-point conformal blocks. These t-channel blocks will be
denoted by  ${}^t{\cal F}^{1234}_l$ (see figure 1b). The
s-t duality of CFT implies that these basis are related,
\ie\ that one can write:
\beq
{}^s{\cal F}^{1234}_p\,\,=\,\,
\sum_{l}\,\,F_{j_p,j_l}\Bigl[\,{2\,\,3\atop
1\,\,4}\Bigr]\,\, {}^t{\cal F}^{1234}_l\,\,,
\label{vnueve}
\eeq
where the $F\Bigl[\,{2\,\,3\atop 1\,\,4}\Bigr]$ matrix is
the so-called fusion matrix \cite{MS,Review}. 

In order to obtain the explicit form of the duality 
transformations of eqs. (\ref{vsiete}), (\ref{vocho}) and
(\ref{vnueve}), we shall make use of a free field
realization of the osp$(1\vert 2)$ current algebra.
This realization was introduced in ref. \cite{ber} and used
in ref. \cite{osp} to study the conformal blocks of the
model. Let us describe briefly its basic features. The field
content of this representation consists of an scalar field
$\phi$, a pair of two conjugate bosonic fields $(w, \chi)$
and two fermionic fields $(\psi, \bar\psi)$ whose
non-vanishing operator product expansions are:
\beq
w(z_1)\,\chi(z_2)\,=\,\psi(z_1)\,\bar\psi(z_2)\,=\,{1\over
z_1-z_2} \,\,\,\,\,\,\,\,\,\,\,\,\,\,\,\,\,\,
\phi(z_1)\,\phi(z_2)\,=\,-{\rm log}\,(z_1-z_2)\,.
\label{treinta}
\eeq
In terms of these fields, one can represent \cite{osp}
easily the primary fields of the model as:
\beq
\Phi^j_m\,=\,\cases{\chi^{j-m}\,e^{-2ij\alpha_0\,\phi}
                     &if $j-m\in \ZZ$\cr\cr
                    \chi^{j-m-{1\over 2}}\,\psi\,
                    e^{-2ij\alpha_0\,\phi}
                    &if $j-m\in \ZZ\,+{1\over 2}\,$,}
\label{tuno}
\eeq
where $\alpha_0\,=\,-1/\sqrt{2k+3}$. As it
is discussed in ref. \cite{osp}, the representation
(\ref{tuno}) is not unique. In fact, there also exists a
conjugate representation which, for a highest weight field,
takes the form:
\beq
\tilde \Phi_j^j\,=\,w^{2j+s}\,\,
e^{2i(j+s)\alpha_0\,\phi}\,,
\label{tdos}
\eeq
where $s\,=\,-k-1$. 
Within this free field approach, the conformal blocks of
the model are computed as vacuum expectation values of
products of fields of the type (\ref{tuno}) and
(\ref{tdos}) and an screening charge $Q$, whose expression
is:
\beq
Q\,=\,\oint\,dz\,
(\,\bar\psi(z)\,-\,w(z)\psi(z)\,)\,e^{i\alpha_0\phi(z)}\,.
\label{ttres}
\eeq
A general four-point block is obtained from a correlator of
the form:
\beq
<\Phi^{j_1}_{m_1}(z_1)\,\Phi^{j_2}_{m_2}(z_2)\,
\Phi^{j_3}_{m_3}(z_3)\,\tilde\Phi^{j_4}_{m_4}(z_4)\,
Q^n\,>\,,
\label{tcuatro}
\eeq
where \cite{osp} the number of screening charges is 
$n\,=\,2\,(\,j_1\,+\,j_2\,+\,j_3\,-\,j_4\,)$. The
basic information about the duality behaviour of the model
can be obtained by studying the block  with $j_3=j_2$ and
$j_4=j_1$ and when the primary fields entering the block
are either highest or lowest weights. Due to this fact, we
shall restrict ourselves to the situation in which
$m_4=-m_1=j_1$ and $m_2=-m_3=j_2$. Moreover, by using the
$sl(2)$ projective invariance of the Virasoro algebra, we
can fix the positions of the four fields to the values
$z_1=0$,
$z_2=z$, $z_3=1$ and $z_4=\infty$. We shall suppose,
finally, that the four representations involved in the
correlator (\ref{tcuatro}) are even. Therefore, according
to these considerations, we have to  study the function: 
\beq
{\cal F}^{1234}(z)\,\equiv\,
<\,\Phi^{j_1}_{-j_1}(0)\,\Phi^{j_2}_{j_2}(z)\,
\Phi^{j_2}_{-j_2}(1)\,\tilde\Phi^{j_1}_{j_1}(\infty)\,
Q^{4j_2}\,>\,.
\label{tcinco}
\eeq
In the framework of this free field representation, the
correlator (\ref{tcinco}) can be given as a multiple
contour integral of the type:
\beq
{\cal F}^{1234}(z)\,=\,
\prod_{i=1}^{n}\,\,\oint_{C_i}\,\,d\tau_i\,
\lambda(z,\{\tau_i\})\,\eta(\{\tau_i\})\,.
\label{tseis}
\eeq
In eq. (\ref{tseis}) (and in what follows), the number $n$ of
integrations is $n=4j_2$. The function 
$\lambda(z,\{\tau_i\})$ in (\ref{tseis}) is the
contribution of the field $\phi$ 
to the correlator (\ref{tcinco}). Taking eqs. (\ref{tuno})
and (\ref{tdos}) into account, one can write this
contribution as:
\bear
\lambda(z,\{\tau_i\})\,=\,
<\,e^{-2ij_1\alpha_0\,\phi(0)}&e^{-2ij_2\alpha_0\,\phi(z)}\,
e^{-2ij_2\alpha_0\,\phi(1)}\,
e^{2i(s+j_1)\alpha_0\,\phi(\infty)}\,\times\rc
&\times e^{i\alpha_0\,\phi(\tau_1)}\cdots
 e^{i\alpha_0\,\phi(\tau_n)}\,>\,.\label{tsiete}\\
\nonumber
\eear
The function $\eta(\{\tau_i\})$ in (\ref{tseis})  represents
the contribution of the fields $w$, $\chi$, $\psi$ and
$\bar\psi$ to  (\ref{tcinco}). Using the Fock space
selection rules of  the free field realization, it was
found in ref. \cite{osp} that this function can be written
as:
\bear
\eta(\{\tau_i\})\,&=&\,(-1)^{2j_2}\,
<\,(\chi(0))^{2j_1}\,(\chi(1))^{2j_2}\,
(w(\infty))^{2j_1+s}\,w(\tau_1)\,\cdots\,
w(\tau_{2j_2})\,\,>\times\rc\rc
&&\times\,<\,\psi(\tau_1)\cdots\psi(\tau_{2j_2})\,
\bar\psi(\tau_{2j_2+1})\cdots\bar\psi(\tau_{4j_2})\,>\,+\,
{\rm permutations.}\label{tocho}\\
\nonumber
\eear

\begin{figure}
\centerline{\hskip-.4in \epsffile{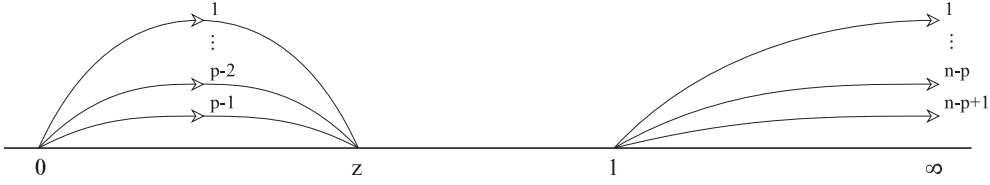}}
\caption{Contours of integration needed to represent 
${}^s{\cal F}^{1234}_p(z)$.}
\label{fig2}
\end{figure}

So far we have not specified the contours appearing in eq.
(\ref{tseis}). This specification is equivalent to the
choice of a  basis in the space of conformal blocks.
The contours corresponding to  well defined s-channel
intermediate states for the blocks (\ref{tseis}) 
have been represented in figure 2. We shall take the first
$n-p+1$ integrals (whose integration variable will be
denoted by $\tau_i=u_i$ for $i=1,\cdots, n-p+1$) along the 
contour joining the points $\tau=1$ and $\tau=\infty$ and
lying on the real axis. The remaining integration variables
will be denoted by $v_i$ (\ie\  $v_i=\tau_{n-p+1+i}$ for
$i=1,\cdots, p-1$ ) and will be integrated in the interval
$(0,z)$. All the integrations in a given interval are
considered as ordered with respect to the location on the
real axis of the singular points of the block (\ie\ $\tau=0,
z, 1$ and $\infty$ ). The ordering along the real line
of these singular points is determined 
by the ordering of the
fields in the correlator (\ref{tcinco}). Thus, for example,
the first field from the left in eq. (\ref{tcinco}) is
evaluated at
$\tau=0$, which is also the first point from the left in
figure 2. Let us denote by 
$\lambda_p(z,\{u_i\},\{v_i\})$ and
$\eta_p(\{u_i\},\{v_i\})$ the functions
$\lambda(z,\{\tau_i\})$ and $\eta(\{\tau_i\})$ after
the relabelling of variables described above. The 
$p^{{\rm th}}$ s-channel block can be represented as: 
\bear
{}^s{\cal F}^{1234}_p(z)\,=\,
\int_1^{\infty}\,du_1\cdots\int_1^{u_{n-p}}\,
du_{n-p+1}\int_0^{z}\,dv_1\cdots\int_0^{v_{p-2}}\,dv_{p-1}\,
\lambda_p(z,\{u_i\},\{v_i\})\,\eta_p(\{u_i\},\{v_i\}).\rc
\label{tnueve}
\eear
By using Wick's Theorem, one can readily evaluate 
$\lambda_p(z,\{u_i\},\{v_i\})$ with the result:
\bear
\lambda_p(z,\{u_i\},\{v_i\})&=&z^{8j_1j_2\rho}\,
(1-z)^{8j_2^2\rho}\,
\prod_{i=1}^{n-p+1}\,u_i^a\,(u_i-z)^b\,(u_i-1)^b\,
\prod_{i<j}(u_i-u_j)^{2\rho}
\times\rc
&&\times
\prod_{i=1}^{p-1}\,v_i^a\,(z-v_i)^b\,(1-v_i)^b
\prod_{i<j}(v_i-v_j)^{2\rho}
\,\prod_{i=1}^{n-p+1}\,\prod_{j=1}^{p-1}
(u_i-v_j)^{2\rho},\rc
\label{cuarenta}
\eear
where $\rho\,=\,\alpha_0^2/2\,=\,1/2(2k+3)$ 
and the constants $a$ and $b$ are defined as 
$a\,=\,-2j_1\alpha_0^2\,$ and  $b\,=\,-2j_2\alpha_0^2\,$.
Notice that the phases chosen in the different powers in 
(\ref{cuarenta}) correspond to the ordering of contours
shown in figure 2. The s-channel intermediate isospin $j_p$,
which  corresponds to the $p^{{\rm th}}$ block, can be
easily obtained \cite{osp} by looking at the $z\rightarrow 0$
behaviour of the function (\ref{tnueve}). One gets:
\beq
j_p\,=\,j_1\,+\,j_2\,+{1\,-\,p\over 2}\,.
\label{cuno}
\eeq

It is now possible to study the implementation of the
different exchange operations defined at the beginning of
this section. Let us, first of all, consider the $\pi_{12}$
operation introduced in eq. (\ref{vseis}). It is clear
that, as the result of acting with  $\pi_{12}$, the blocks 
(\ref{tcinco}) are transformed into: 
\beq
{\cal F}^{2134}(z)\,\equiv\,
<\,\Phi^{j_2}_{j_2}(z)\,\Phi^{j_1}_{-j_1}(0)\,
\Phi^{j_2}_{-j_2}(1)\,\tilde\Phi^{j_1}_{j_1}(\infty)\,
Q^{4j_2}\,>\,.
\label{cdos}
\eeq
As in eq.
(\ref{tnueve}), one can get an integral  representation of
the blocks  (\ref{cdos}) for a well-defined s-channel
intermediate state. Notice that now the contour ordering
corresponding to the correlator (\ref{cdos}) is the one shown
in figure 3. Therefore we can write:
\beq
{}^s{\cal F}^{2134}_p(z)\,=\,
\int_1^{\infty}\,du_1\cdots\int_1^{u_{n-p}}\,
du_{n-p+1}
\int_z^{0}\,d\bar v_1\cdots\int_z^{\bar v_{p-2}}
\,d\bar v_{p-1}\,\bar \lambda_p(z,\{u_i\},\{\bar v_i\})\,
\eta_p(\{u_i\},\{\bar v_i\}),
\label{ctres}
\eeq
where the function $\eta_p$ is the same used in eq.
(\ref{tnueve}) and $\bar \lambda_p(z,\{u_i\},\{\bar v_i\})$
is given by:
\bear
\bar \lambda_p(z,\{u_i\},\{\bar v_i\})
&=&(-z)^{8j_1j_2\rho}\, (1-z)^{8j_2^2\rho}\,
\prod_{i=1}^{n-p+1}\,u_i^a\,(u_i-z)^b\,(u_i-1)^b\,
\prod_{i<j}(u_i-u_j)^{2\rho}
\times\rc
&&\times
\prod_{i=1}^{p-1}\,(-\bar v_i)^a\,(\bar v_i-z)^b
\,(1-\bar v_i)^b
\prod_{i<j}(\bar v_i-\bar v_j)^{2\rho}
\,\prod_{i=1}^{n-p+1}\,\prod_{j=1}^{p-1}
(u_i-\bar v_j)^{2\rho},\rc
\label{ccuatro}
\eear
\begin{figure}
\centerline{\hskip-.4in \epsffile{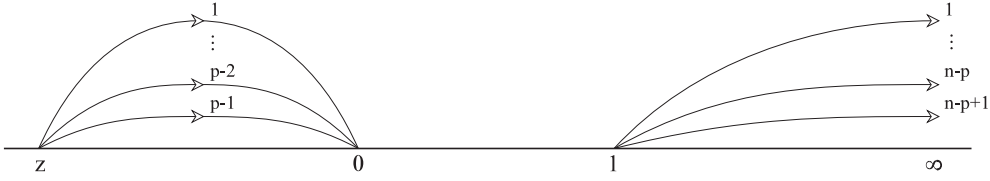}}
\caption{Contours of integration used in the free field
representation of ${}^s{\cal F}^{2134}_p(z)$.}
\label{fig3}
\end{figure}

Let us now try to relate the functions (\ref{tnueve}) and 
(\ref{ctres}). First of all, it is clear from the contours
of figure 3 that the integral (\ref{ctres}) is naturally
defined in the situation in which $z<0$, whereas, on the
contrary, the integration limits in (\ref{tnueve})
correspond to the case $z>0$. It is not difficult to
analytically continue the expression (\ref{ctres}) from the 
$z<0$ domain to the range $z>0$. The first step in this
analytical continuation consists in exchanging the upper
and lower limits of the $\bar v_i$ integrations in 
(\ref{ctres}):
\beq
\int_z^{0}\,d\bar v_1\cdots
\int_z^{\bar v_{p-2}}\,d\bar v_{p-1}
\,[\,\cdots\,]\,=\,(-1)^{p-1}\,
\int^z_{0}\,d\bar v_1\cdots
\int^z_{\bar v_{p-2}}\,d\bar v_{p-1}
\,[\,\cdots\,]\,\,.
\label{ccinco}
\eeq
Notice that, in the right-hand side of eq. (\ref{ccinco}),
the variables are ordered as 
$z>\bar v_{p-1}>\bar v_{p-2}>\cdots >\bar v_{1}>0$. This is
not the ordering appearing in the integral (\ref{tnueve}).
The latter can be obtained by means of the following
redefinition of the $\bar v_i$ variables:
\beq
v_i\,=\,\bar v_{p-i}
\,\,\,\,\,\,\,\,\,\,\,\,\,\,\,
i=1,\cdots, p-1\,\,.
\label{cseis}
\eeq
Performing this change of variables in the right-hand side
of eq. (\ref{ccinco}) and reversing the order of the
iterated integrals in the resulting expression, one gets:
\beq
\int_z^{0}\,d\bar v_1\cdots\int_z^{\bar v_{p-2}}\,d\bar v_{p-1}
\,[\,\cdots\,]\,=\,(-1)^{p-1}\,
\int_0^{z}\,dv_1\cdots\int_0^{v_{p-2}}\,dv_{p-1}\,
\,[\,\cdots\,]\,\,.
\label{csiete}
\eeq
Moreover, it is clear from their definitions that, when the
relabelling (\ref{cseis}) is performed, the function 
$\bar \lambda_p(z,\{u_i\},\{\bar v_i\})$ is transformed
into the function $\lambda_p(z,\{u_i\},\{v_i\})$ multiplied
by a phase, while the function $\eta_p$ is multiplied by a
sign. It is easy to evaluate these factors. The result is:
\bear
\bar \lambda_p(z,\{u_i\},\{\bar v_i\})
&=&e^{8i\pi j_1j_2\rho}\,e^{i\pi (a+b)(p-1)}\,
e^{i\pi (p-1)(p-2)\rho}\,
\lambda_p(z,\{u_i\},\{v_i\})\rc
\eta_p(\{u_i\},\{\bar v_i\})&=&
(-1)^{{(p-1)(p-2)\over 2}}\,
\eta_p(\{u_i\},\{ v_i\})\,\,.\label{cocho}\\
\nonumber
\eear
Putting together  all the factors appearing in  eqs.
(\ref{csiete}) and  (\ref{cocho}), we obtain the explicit
expression of $\Lambda_{1,2}^p$:
\beq
\Lambda^{p}_{1,2}\,=\,(-1)^{{p(p-1)\over 2}}\,
e^{i\pi\,(h_{j_p}\,-\,h_{j_1}\,-\,h_{j_2})}\,\,,
\label{cnueve}
\eeq
where $h_{j_1}$, $h_{j_2}$ and $h_{j_p}$ are given by eq. 
(\ref{cinco}). It is interesting to write the sign in eq.
(\ref{cnueve}) in a slightly different form. Let us denote
by $<x>$ the integer part of any integer or half-integer
number $x$. As $(-1)^{p(p-1)/2}\,=\,(-1)^{<p/2>}$ for
any $p\in\ZZ$, and after taking eq. (\ref{cuno}) into
account, eq. (\ref{cnueve}) can be rewritten as:
\beq
\Lambda^{p}_{1,2}\,=\,(-1)^{<j_1+j_2-j_p+{1\over 2}>}\,
e^{i\pi\,(h_{j_p}\,-\,h_{j_1}\,-\,h_{j_2})}\,\,.
\label{cincuenta}
\eeq
It is also very interesting 
to write eq. (\ref{cincuenta}) in
terms of the deformation parameter $q$, introduced in eq. 
(\ref{dseis}). After a short calculation, one concludes
that the corresponding expression is:
\beq
\Lambda^{p}_{1,2}\,=\,(-1)^{<j_1+j_2-j_p+{1\over 2}>}\,
q^{j_p(2j_p+1)\,-\,j_1(2j_1+1)\,-\,j_2(2j_2+1)}\,\,.
\label{ciuno}
\eeq
In the form (\ref{ciuno}), the value of $\Lambda^{p}_{1,2}$
has a very neat interpretation. Indeed, one can regard the
sign in (\ref{ciuno}) as the classical part 
of $\Lambda^{p}_{1,2}$ and the power
of $q$ as its quantum deformation. The quantum part of 
$\Lambda^{p}_{1,2}$ is the one expected from the general
formalism of CFT \cite{MS}, whereas the classical
contribution should be determined from the (undeformed) 
${\rm osp}(1\vert 2)$ representation theory. In general,
the state $|\,j_p,m_p>$, obtained by tensor
multiplication of two representations of isospins $j_1$ and
$j_2$, is given by an expression of the form:
\beq
|\,j_p,m_p\,>\,=\,\sum_{m_1,m_2}\,
C_{j_1,m_1, \lambda_1;j_2,m_2,\lambda_2}
^{j_p,m_p,\lambda_p}\,\,\,\,
|j_1,m_1>\otimes\,|j_2,m_2>\,\,,
\label{cidos}
\eeq
where $C_{j_1,m_1, \lambda_1;j_2,m_2,\lambda_2}
^{j_p,m_p,\lambda_p}$ are the ${\rm osp}(1\vert 2)$
Clebsch-Gordan coefficients. Under the exchange of the two
representations in the tensor product, these 
Clebsch-Gordan coefficients change by a sign. Let us write
this behaviour as:
\beq
C_{j_2,m_2, \lambda_2;j_1,m_1,\lambda_1}
^{j_p,m_p,\lambda_p}\,\,=\,\,
\epsilon_{j_1,m_1, \lambda_1;j_2,m_2,\lambda_2}
^{j_p,m_p,\lambda_p}\,\,\,\,
C_{j_1,m_1, \lambda_1;j_2,m_2,\lambda_2}
^{j_p,m_p,\lambda_p}\,\,.
\label{citres}
\eeq
The $\epsilon_{j_1,m_1, \lambda_1;j_2,m_2,\lambda_2}
^{j_p,m_p,\lambda_p}$ signs of the right-hand side of 
(\ref{citres}) can be obtained from the values of the 
Clebsch-Gordan coefficients of ${\rm osp}(1\vert 2)$
\cite{Pais}. One has:
\bear
\epsilon_{j_1,m_1, \lambda_1;j_2,m_2,\lambda_2}
^{j_p,m_p,\lambda_p}\,\,=\,\,
\,(-1)^{<j_1+j_2-j_p+{1\over 2}>}
(-1)^{(\lambda_p+1)(\lambda_1+\lambda_2)+\lambda_1\lambda_2}
\,(-1)^{(\lambda_1+2(j_1-m_1))(\lambda_2+2(j_2-m_2))}\,\,.
\rc
\label{cicuatro}
\eear
Notice that, indeed, when $\lambda_1\,=\,\lambda_2\,=\,0$
and $j_1-m_1\,,\,j_2-m_2\in\ZZ$, the signs of the right-hand
sides of (\ref{cicuatro}) and (\ref{ciuno}) coincide.

It is
interesting to point out the dependence on the Cartan
eigenvalues $m_i$ of the right-hand side of eq.
(\ref{cicuatro}). This dependence, which does not occur in
the su$(2)$ case, has its origin in the sign generated in
the exchange of two ${\rm osp}(1\vert 2)$ states due to
their different Grassmann parities. It follows that, in
general, the eigenvalues in eq. (\ref{vsiete}) will depend
on the $m_i$'s. It can be checked that, for general Cartan
eigenvalues of the fields inserted in the four-point
correlator, our free field representation gives rise to the
same $m_i$ dependence as in eq. (\ref{cicuatro}).

The behaviour of the blocks under the exchange operation
$\pi_{34}$ can be determined by the same method employed to
study the action of $\pi_{12}$. In fact, it is easy to
verify that, with the appropriate substitutions, the value
of the constants $\Lambda_{3,4}^p$ is also given by
eq. (\ref{cincuenta}). The determination of the braiding
matrix defined in eq. (\ref{vocho}) is much more involved.
In general, one has to employ contour manipulation
techniques in order to relate the multiple integrals
appearing in the free field representation of both sides of
eq. (\ref{vocho}). We shall restrict ourselves here to
analyze the braiding matrix of blocks of the type
(\ref{tcinco}) with $j_1=j_2=j$. Let us denote by 
${\cal B}^{{j}}$ the 
$(4j+1)\times (4j+1)$ dimensional matrix
that implements the s-u crossing symmetry in the free field
representation of this type of blocks. In appendix A, 
the $j=1/2$ case is worked out. Using the results of this
appendix one can write the ${\cal B}^{{1\over 2}}$ matrix as:

\bear
{\cal B}^{{1\over 2}}\,=\,
\pmatrix{{q^2\over [{3\over 2}]_+}&
-q &-q^{-1}\,{[{5\over
2}]_+\over [{3\over 2}]_+}\cr\cr
{q\over  [{3\over 2}]_+}&
-{[1]_+ +[2]_+\over [2]_+}&
 -q^{-2}\,{[{5\over 2}]_+[1]_+
\over [2]_+[{3\over 2}]_+}\cr\cr
-{q^{-1}\over[{3\over 2}]_+}&
{q^{-2}\,[1]_+\over [2]_+}
& -q^{-4}\,{[1]_+\over
[2]_+[{3\over 2}]_+}\cr}\,\,,
\label{cicinco}\\
\nonumber
\eear
where the graded q-numbers have been defined in eq.
(\ref{dsiete}). As a consistency check of eq. 
(\ref{cicinco}) one can verify that, as expected, when
$j_1=j_2=1/2$, the quantities $\Lambda_{1,2}^p$, given in
eq. (\ref{cincuenta}), are eigenvalues of 
${\cal B}^{{1\over 2}}$. Moreover, it is interesting to
point out that, performing a change of basis in the space
of  blocks, the braiding matrix of eq. (\ref{cicinco}) can
be recast as a symmetric matrix which is much more
convenient for our purposes. Indeed, let us conjugate the
braiding matrix ${\cal B}^{{1\over 2}}$ in the form:
\beq
B^{{1\over 2}}\,=\,\gamma\,
{\cal B}^{{1\over 2}}\,\gamma^{-1}\,\,,
\label{ciseis}
\eeq
where $\gamma$ is the following diagonal matrix:
\beq
\gamma\,=\,
\pmatrix{1
&&\cr
&i\sqrt{\Bigl[\,{3\over 2}\,\Bigr]_{+}}&\cr
&&\sqrt{\Bigl[\,{5\over 2}\,\Bigr]_{+}}
\,\,\cr}\,\,.
\label{cisiete}
\eeq
After the conjugation (\ref{ciseis}), the resulting
braiding matrix $B^{{1\over 2}}$ is:

\bear
B^{{1\over 2}}\,=\,
\pmatrix{{q^2\over [{3\over 2}]_+}&
i{q\over \sqrt{ [{3\over 2}]_+}}
&-q^{-1}{\sqrt{[{5\over 2}]_+}\over
[{3\over 2}]_+}\cr\cr
i{q\over \sqrt{ [{3\over 2}]_+}}&
-{[1]_+ +[2]_+\over [2]_+}&
 -iq^{-2}\,{[1]_+\sqrt{[{5\over 2}]_+}
\over [2]_+\sqrt{[{3\over 2}]_+}}\cr\cr
-q^{-1}{\sqrt{[{5\over 2}]_+}\over
[{3\over 2}]_+}&
 -iq^{-2}\,{[1]_+\sqrt{[{5\over 2}]_+}
\over [2]_+\sqrt{[{3\over 2}]_+}}
& -q^{-4}\,{[1]_+\over
[2]_+[{3\over 2}]_+}\cr}\,\,.
\label{ciocho}\\
\nonumber
\eear
On the other hand, as  argued in ref. \cite{DF}, the fusion
matrix for the blocks (\ref{tcinco}) can be obtained by
looking at the relation between the functions 
${}^s{\cal F}^{1234}(z)$ and ${}^s{\cal F}^{1234}(1-z)$. In
the simplest case, in which $j_1=j_2=1/2$, the corresponding
matrix elements have been explicitly given in ref.
\cite{DF}. Adapting eq. (5.11) of the 
first paper in  ref. \cite{DF} to our case, and after
conjugating with the same diagonal matrix
$\gamma$ as in eq. (\ref{cisiete}), one arrives at the
following symmetric expression:

\bear
F^{{1\over 2}}\,=\,
\pmatrix{-{1\over [{3\over 2}]_+}&
-{i\over \sqrt{ [{3\over 2}]_+}}
&-{\sqrt{[{5\over 2}]_+}\over [{3\over 2}]_+}\cr\cr
-{i\over \sqrt{ [{3\over 2}]_+}}&{[1]_+ +[2]_+\over
[2]_+}& -i{[1]_+\sqrt{[{5\over 2}]_+}\over
[2]_+\sqrt{[{3\over 2}]_+}}\cr\cr -{\sqrt{[{5\over
2}]_+}\over [{3\over 2}]_+}& -i{[1]_+\sqrt{[{5\over
2}]_+}\over [2]_+\sqrt{[{3\over 2}]_+}}& {[1]_+\over
[2]_+[{3\over 2}]_+}\cr}\,\,,
\label{cinueve}\\
\nonumber
\eear
where we have denoted by $F^{j}$ the fusion matrix for the
blocks (\ref{tcinco}) with $j_1=j_2=j$.

From the particular cases of $B^{j}$ and $F^{j}$ just found
it is not difficult to find out their values for an
arbitrary value of the isospin $j$. The key observation in
this respect is the comparison between the matrices of eqs. 
(\ref{ciocho}) and (\ref{cinueve}) and those corresponding
to the su$(2)$ CFT. Let us denote by $\tilde B^{j}(t)$ and
$\tilde F^{j}(t)$ the braiding and fusion matrices of
the su$(2)$ conformal blocks of correlators of four primary
fields with the same isospin $j$.  For convenience, we have
chosen a notation for these matrices in which the
deformation parameter $t$ appears explicitly. As was shown
in ref. \cite{AGS}, the su$(2)$ fusion matrices can be put in
terms of the Racah-Wigner $6j$ symbols of $U_t({\rm
su}(2))$. These symbols were computed in ref. \cite{KR}.
Using these results we can write:
\bear
\tilde F_{j_1\,j_2}^{\,j}(t)\,&=&\,
(-1)^{j_1+j_2-2j}\,\,\sqrt{[2j_1+1]}\,
\sqrt{[2j_2+1]}\,\,
\Bigl(\,\Delta(j,j,j_1)\,\Delta(j,j,j_2)\,
\Bigr)^{2}\,\times\rc
&&\times\sum_{m\ge 0}\,\,(-1)^{m}
\,[m+1]\,!\,\,\Bigl(\,[m-2j-j_1]\,!\,\Bigr)^{-2}\,\,
\Bigl(\,[m-2j-j_2]\,!\,\Bigr)^{-2}\,\,\times\rc
&&\times\,\,\Bigl(\,[2j+j_1+j_2-m]\,!\,\Bigr)^{-2}\,\,
\Bigl(\,[4j-m]\,!\,\Bigr)^{-1}\,\,.\label{sesenta}\\
\nonumber
\eear
The q-numbers appearing in the right-hand side of eq. 
(\ref{sesenta}) have been defined in eq. (\ref{vtres}). The
function $\Delta\,(a,b,c)$ is defined as:
\beq
\Delta\,(a,b,c)\,=\,\sqrt{{[-a+b+c]\,!\,\,
[a-b+c]\,!\,\,[a+b-c]\,!\over
[a+b+c+1]\,!}}\,\,.
\label{suno}
\eeq
Let us now argue that the same identification  used in
section 2 to pass from the su$(2)$ to the 
${\rm osp}(1\vert 2)$ quantum numbers (\ie\ $t=-q$) can be
utilized to connect the fusion matrices. Actually, our claim
is that:
\beq
F_{j_1\,j_2}^{j}\,=\,
\tilde F_{2j_1\,2j_2}^{\,2j}(-q)\,\,.
\label{sdos}
\eeq
In order to evaluate the right-hand side of eq.
(\ref{sdos}), we shall use the fact that, when $t=-q$, the
quantum factorials are related as:
\beq
[x]\,!\,=\,(-1)^{x(x-1)\over 4}\,\,
\Bigl[\,{x\over 2}\,\Bigr]_{+}\,!\,\,,
\label{stres}
\eeq
where we have denoted:
\beq
[y]_+\,!\,=\,\prod_{j=1/2\atop 2j\,\in\,\ZZ}^{y}\,\,
[\,j\,]_+\,\,.
\label{scuatro}
\eeq
In terms of the factorials (\ref{scuatro}), we define the
function $\Delta_{+}\,(a,b,c)$ by means of the expression:
\beq
\Delta_{+}\,(a,b,c)\,=\,\sqrt{{[-a+b+c]_+\,!\,\,
[a-b+c]_+\,!\,\,[a+b-c]_+\,!\over
[a+b+c+{1\over 2}]_+\,!}}\,\,.
\label{scinco}
\eeq
Using these definitions, eq. (\ref{sdos}) can be written as:
\bear
F_{j_1\,j_2}^{j}\,&=&\,
(-1)^{2(j_1+j_2)^2\,-\,j_1\,-\,j_2}\,
i^{\,2(j_1-<j_1>)}\,i^{\,2(j_2-<j_2>)}\,
\,\,\sqrt{\Bigl[\,{4j_1+1\over 2}\Bigr]_+}\,\,
\sqrt{\Bigl[\,{4j_2+1\over 2}\Bigr]_+}\,\times\rc\rc
&&\times\,\Bigl(\,\Delta_{+}(j,j,j_1)\,
\Delta_{+}(j,j,j_2)\,\Bigr)^{2}\,
\sum_{{m\ge 0\atop 2m\in\ZZ}}\,\,(-1)^{m(2m-1)}
\,\Bigl[{2m+1\over 2}\Bigr]_+\,!\,\,
\Bigl(\,[m-2j-j_1]_+\,!\,\Bigr)^{-2}\,\times\rc\rc
&&\times\Bigl(\,[m-2j-j_2]_+\,!\,\Bigr)^{-2}\,
\,\Bigl(\,[2j+j_1+j_2-m]_+\,!\,\Bigr)^{-2}\,\,
\Bigl(\,[4j-m]_+\,!\,\Bigr)^{-1}\,\,.\rc
\label{sseis}
\eear
Several remarks concerning eq. (\ref{sseis}) are in order. 
First of all, one must be specially careful with the 
$t=-q$ identification in the square root terms of the
right-hand side of eq. (\ref{sesenta}), since minus signs
are generated inside the square root and one must give a
prescription to deal with them. To obtain eq. (\ref{sseis}),
we have taken $\sqrt{(-1)^{2x}}$ for $2x\in\ZZ$ to be
$i^{2(x-<x>)}$. Moreover, when $F_{j_1\,j_2}^{j}$ is given
by eq. (\ref{sseis}), one can prove a set of properties
satisfied by the fusion matrix. The most interesting for
our purposes have been compiled in appendix B. 

Let us now provide evidence supporting our claim of eq. 
(\ref{sdos}). Our first argument is the fact that a direct
substitution for $j=1/2$ shows that the matrix elements
computed from eq. (\ref{sseis}) coincide with the ones
displayed in eq. (\ref{cinueve}). Another piece of evidence
is obtained by computing the braiding matrix from eq. 
(\ref{sseis}). Indeed, a general argument in CFT allows to
relate the fusion and braiding matrices \cite{MS}. This
relation involves the signs (\ref{cicuatro}), which encode
the behaviour of the Clebsch-Gordan coefficients under the
permutation of the two representations that are multiplied
in the tensor product. In our case, the relation between
the braiding and fusion matrices takes the form:
\beq
B_{j_pj_l}\Bigl[\,{2\,\,3\atop 1\,\,4}\Bigr]\,=\,
\epsilon_{j_3,m_3, \lambda_3;j_4,m_4,\lambda_4}
^{j_p,m_p,\lambda_p}\,\,\,\,
\epsilon_{j_l,m_l, \lambda_l;j_3,m_3,\lambda_3}
^{j_1,m_1,\lambda_1}\,\,\,\,
e^{i\pi(h_{j_1}+h_{j_4}-h_{j_p}-h_{j_l})}\,\,\,
F_{j_pj_l}\Bigl[\,{2\,\,4\atop 1\,\,3}\Bigr]\,\,.
\label{ssiete}
\eeq
Taking $j_1=j_2=j_3=j_4=j$ and using eq. (\ref{cicuatro})
in eq. (\ref{ssiete}), one can prove that:
\beq
B_{j_pj_l}^{j}\,=\,(-1)^{2j}\,(-1)^{<j_p>}\,(-1)^{<j_l>}\,\,
q^{4c_j-2c_{j_p}-2c_{j_l}}\,\,F_{j_pj_l}^{j}\,\,,
\label{socho}
\eeq
from which we can obtain the general form of $B^j$ once
$F^j$ is known. One can easily check that, for $j=1/2$,  
eq. (\ref{socho}) reproduces the
expression of the braiding matrix elements given in eq. 
(\ref{ciocho}). For a general value of $j$
one can verify that the eigenvalues of the matrix 
(\ref{socho})  coincide with our free field expression (\ie\
with eq. (\ref{ciuno}) for $j_1=j_2=j$). For low values of
$j$ this statement can be checked by an explicit
calculation. There is, however, a more powerful indirect
argument that makes use of the relation between the matrix 
$B^j$ and its counterpart $\tilde B^{\,j}(t)$ in the su$(2)$
theory. The matrix elements of the latter are given in terms
of those of the su$(2)$ fusion matrix $\tilde F^{\,j}(t)$
by:
\beq
\tilde B_{j_pj_l}^{\,j}(t)\,=\,(-1)^{j_p+j_l-2j}\,\,
t^{j(j+1)-{j_p(j_p+1)\over 2}
-{j_l(j_l+1)\over 2}}\,\,\tilde F_{j_pj_l}^{\,j}(t)\,\,.
\label{snueve}
\eeq
It is not difficult now to prove the following relation
between the matrices of eqs. (\ref{snueve}) and
(\ref{socho}):
\beq
B_{j_pj_l}^{\,j}\,=\,(-1)^{2j}\,
\tilde B_{2j_p,2j_l}^{\,2j}(-q)\,\,.
\label{setenta}
\eeq
Eq. (\ref{setenta}), which was obtained under the
assumption that eq. (\ref{sseis}) is correct, implies that
the eigenvalues of the matrices $B^j$ and 
$(-1)^{2j}\,\tilde B^{\,2j}(-q)$ are equal. As the
eigenvalues of $\tilde B^{\,j}(t)$, which we shall denote
by $\tilde\Lambda_j^{\,l}(t)$ for $l=0, \cdots, 2j$, are
known, we can obtain in this way the eigenvalues of $B^j$.
We are now going to check that the latter agree with the
values written in eq. (\ref{ciuno}). Indeed, it is well-known
that the  $\tilde \Lambda_j^{\,l}(t)$ are given by:
\beq
\tilde \Lambda_{j}^{\,l}(t)\,=\,(-1)^{2j-l}\,
t^{{l(l+1)\over 2}-j(j+1)}\,\,.
\label{stuno}
\eeq
The eigenvalues of $(-1)^{2j}\,\tilde B^{\,2j}(-q)$ are
$(-1)^{2j}\,\tilde \Lambda_{2j}^{\,2l}(-q)$. A 
straightforward calculation shows that:
\beq
(-1)^{2j}\,\tilde \Lambda_{2j}^{\,2l}(-q)\,=\,
(-1)^{<2j-l+{1\over 2}>}\,q^{l(2l+1)-2j(2j+1)}\,\,,
\label{stdos}
\eeq
which, for $l=0, \cdots, 2j$ and $2l\in \ZZ$,  coincide with
the set of values given in eq. (\ref{ciuno}). This proves
our statement.

\setcounter{equation}{0}
\section{osp$\bf{(1\vert 2)}$ invariants for torus
knots and links }

In the previous section we have characterized the exchange
symmetry of the ${\rm osp}(1\vert 2)$ CFT. It is nowadays an
established fact that there exists a non-trivial connection
between the duality properties of two-dimensional CFT's and
three-dimensional topology. The best way to uncover this
connection is by formulating a suitable Chern-Simons (CS)
topological field theory in three dimensions \cite{witten}. 
This CS theory must be such that, after quantization, its
states are in one-to-one correspondence with the conformal
blocks of the two-dimensional CFT. In our case there is an
obvious choice for the three-dimensional theory. Indeed,
since the CFT we are dealing with is endowed of an  ${\rm
osp}(1\vert 2)$ current algebra, it is natural to consider a
theory based on the action:
\beq
S={k\over4\pi}\int_M {\rm Str} \Big [ A\wedge dA 
+{2\over 3}A\wedge A\wedge A  \Big], 
\label{sttres}
\eeq
where $A$ is a one-form connection taking values in the 
${\rm osp}(1\vert 2)$ superalgebra and $M$ is a
three-dimensional manifold without boundary. In eq. 
(\ref{sttres}), $k$ is a non-negative integer. Once the
connection with the two-dimensional theory  be 
established, $k$ will be identified with the level of the 
${\rm osp}(1\vert 2)$ affine symmetry.
\begin{figure}
\centerline{\hskip.4in \epsffile{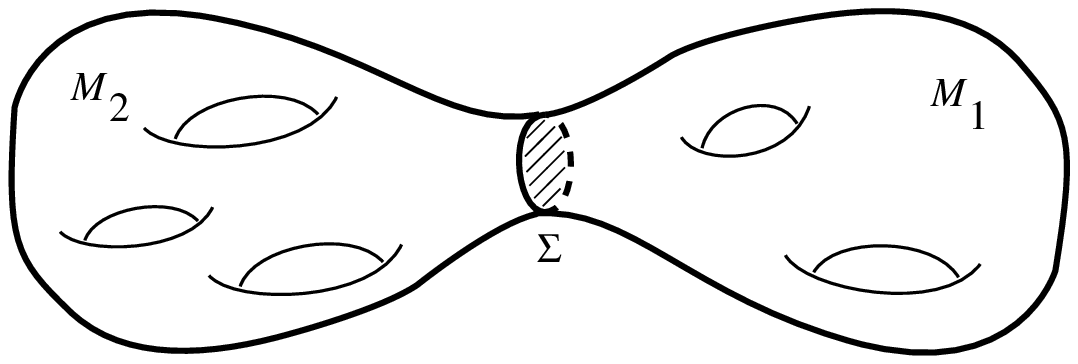}}
\caption{The three dimensional manifold $M$ is split into
$M_1$ and $M_2$, joined  along their common boundary
$\Sigma$}
\label{fig4}
\end{figure}

The basic observables in the CS theories are the Wilson
line operators. These are operators defined for each closed 
curve in $M$ and for each irreducible
representation of the superalgebra. For an isospin $j$
representation, the Wilson line operator for a curve
$\gamma$ is given by:
\beq
W_j^{\gamma}\,\equiv\,{\rm Str_j}\,\Bigl[\,
P\,{\rm exp}\Bigl(\,{\int_{\gamma}\,A}\Bigr)\,\Bigr]\,\,,
\label{stcuatro}
\eeq
where $P$ denotes a path-ordered product along 
$\gamma$, and the supertrace is taken as the trace over the
bosonic states minus the trace over the fermionic states of
the isospin $j$ representation of ${\rm osp}(1\vert 2)$.
Notice that the operators $W_j^{\gamma}$ are both gauge
invariant and metric independent.

In order to quantize the theory based on the action
(\ref{sttres}), one must decompose the manifold $M$ as the
connected sum of two three-manifolds $M_1$ and $M_2$
sharing a common boundary $\Sigma$ (see figure 4). In
general, the identification of the boundaries of $M_1$ and
$M_2$ will be performed through a homeomorphism. The
surface $\Sigma$ in this decomposition will be our
equal-time quantization surface. The topological nature of
the CS theory allows the possibility of choosing different
decompositions of the same three-manifold $M$. The quantum
Hilbert space of states of the theory will depend on these
decompositions. 

The quantization of the CS theory is performed in
the presence of Wilson line operators of the type 
(\ref{stcuatro}). In general, the curves on which these
Wilson lines are defined can intersect with $\Sigma$. In
this case, we shall have on $\Sigma$ a quantization problem
with punctures. Each of these punctures is characterized by
a representation of the gauge group and by the coordinates of
a point of $\Sigma$. According to ref. \cite{witten}, there
exists a correspondence between the CS states on $\Sigma$
and the conformal blocks of a CFT defined on the same
two-dimensional surface. These conformal blocks correspond
to correlators of fields, with the quantum numbers of the
Wilson lines, which are inserted at  the points of
$\Sigma$ where the intersection with the
three-dimensional curves takes place. The interesting
aspect  from the topological point of view is that the vacuum
expectation values of products of Wilson lines are
topological invariant and, therefore, one expects that they
could be related to some link polynomials. The connection
between the CS gauge theory and CFT provides a powerful
method to compute these link invariants. 

In this section we shall consider the case in which the
manifolds $M_1$ and $M_2$ are two solid tori whose boundary
$\Sigma$ is a torus $T^2$. We shall assume that the Wilson
lines do not intersect with the boundary torus. According
to the general arguments reviewed above, the states
associated to this two-dimensional quantization surface
should be in correspondence with the zero-point conformal
blocks of the torus, \ie\ with the supercharacters of the
model. In what follows we shall verify this fact and we
shall find a set of operators which, acting on the torus
states, represent the fusion rules of the 
${\rm osp}(1\vert 2)$ CFT. Within this approach we shall be
able to compute vacuum expectation values of torus links in
the three-sphere $S^3$. In the next section, based on this
result, we shall develop a formalism which will allow us to
compute expectation values  of more general classes of
links.

\begin{figure}
\centerline{\hskip-.8in \epsffile{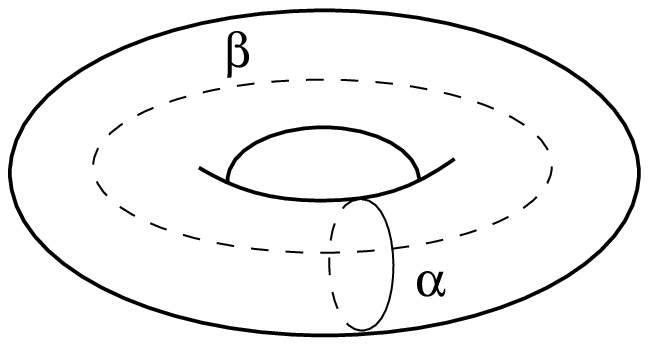}}
\caption{Canonical homology basis for the torus}
\label{fig5}
\end{figure}

When $\Sigma=T^2$ one can argue, as in ref. \cite{LR}, that
the only relevant components of the connection $A$ on the
torus are its zero-modes, which parametrize the holonomy of
the gauge field around the non-trivial homology cycles of
$T^2$. Let us choose a basis for the first homology of the
torus as shown in figure 5, in which the $\alpha$ cycle is
the one which is contractible in the solid torus. The
holomorphic one-form $\omega$ is defined by its integrals
along the $\alpha$ and $\beta$ cycles:
\beq
\int_{\alpha}\,\omega\,=\,1
\,\,\,\,\,\,\,\,\,\,\,\,\,\,
\int_{\beta}\,\omega\,=\,\tau\,\,,
\label{stcinco}
\eeq
where $\tau$ is the modular parameter of the torus. Since
the first homology group of $T^2$ is abelian, we can take
the zero-mode part of $A$ in the Cartan subalgebra of 
${\rm osp}(1\vert 2)$. We shall use the parametrization:
\beq
A\,=\,{\pi a\over \tau_2}\,\bar\omega\,J_0^0\,-\,
{\pi \bar a\over \tau_2}\,\omega\,J_0^0 \,\,,
\label{stseis}
\eeq
where  $\tau_2\,=\,{\rm Im}\,\tau$, $a$ and $\bar a$ are
constants and $J_0^0$ is the ${\rm osp}(1\vert 2)$ Cartan
generator. In the framework of the path integral
quantization of the CS action (\ref{sttres}), one can
formulate \cite{LR} an effective problem for the zero-modes
of the gauge field. One of the outcomes of this formalism is
the fact that, in the effective theory, the coefficient $k$ 
of the CS action is shifted by $c_v$, the quadratic
Casimir in the adjoint representation. For the 
${\rm osp}(1\vert 2)$ superalgebra, $c_v$ is equal to $3/2$
and, therefore, the above-mentioned shift is 
$k\rightarrow k+{3\over 2}$. The states appearing in the
zero-mode problem are functions of the variable $a$, whose
form can be obtained by solving the Gauss law associated to
the action (\ref{sttres}). In
fact, adapting the result of ref. \cite{LR} to our 
${\rm osp}(1\vert 2)$ case, one can readily prove that the
states are given by the numerator of the supercharacters 
(\ref{once}) multiplied by a convenient prefactor. These
functions are:
\beq
\xi_{j,k}(a,\tau)\,\equiv\,
e^{{\pi(2k+3)\over 8\tau_2}\,a^2}\,\,
[\,\,\vartheta_{4j+1,2k+3}(a,\tau)\,-\,
\vartheta_{-4j-1,2k+3}(a,\tau)\,\,]\,\,,
\label{stsiete}
\eeq
where $j$ is integer or half integer. From the periodicity
properties of the characters (eq. (\ref{trece})), one can
immediately conclude that there only exist  $k+1$
independent states in the effective Hilbert space whose
wavefunctions are given by  $\xi_{j,k}(a,\tau)$ for 
$j=0, {1\over 2},\cdots, {k\over 2}$.

It is not difficult to obtain the operator realization of
the gauge field (\ref{stseis}) in the Hilbert space
spanned by the functions (\ref{stsiete}). Actually, the
canonical commutation relations corresponding to the
action (\ref{sttres}), after taking into account the 
$k\rightarrow k+{3\over 2}$ shift, determine the
commutator of the zero-mode components of the gauge field.
This commutator is:
\beq
[\,\bar a\,,\,a\,]\,=\,{4\tau_2\over \pi(2k+3)}\,\,,
\label{stocho}
\eeq
which implies that $\bar a$ can be represented
as:
\beq
\bar a\,=\,{4\tau_2\over \pi(2k+3)}\,
{\partial\over \partial a}\,\,.
\label{stnueve}
\eeq
Using eqs. (\ref{stcinco}), (\ref{stseis}) and 
(\ref{stnueve}) one can easily find the expression of the
Wilson line operators in terms of $a$ and
$\partial/\partial a$. Actually, this can only be done for
Wilson line operators corresponding to torus knots, \ie\ for
curves that can be drawn on the surface of $T^2$ without
self-intersections. Let $\gamma_{r,s}$ be a torus knot on
$T^2$ which belongs to the same homology class as 
$r\alpha+s\beta$, for two coprime integers $r$ and $s$.  We
shall  denote by $W_j^{(r,s)}$ the Wilson line operator 
(\ref{stcuatro}) for $\gamma=\gamma_{r,s}$. If $\Lambda_j$
represents the set of eigenvalues of the Cartan generator 
$J_0^0$  in the isospin $j$ representation, \ie\ 
the $4j+1$ values  
$\Lambda_j\,=\,\{\,j-{p\over 2}\,,\,p=0\,\cdots\,,4j\}$,
the expression of the $W_j^{(r,s)}$ operator is given by:
\beq
W_j^{(r,s)}\,=\,\sum_{n\in\Lambda_j}\,\,
(-1)^{2(j-n)}\,\,{\rm exp}\,
\Bigl[\,{n\pi(r+s\bar\tau)\over \tau_2}\,a\,-\,
{4n(r+s\tau)\over 2k+3}\,{\partial\over\partial a}
\,\Bigr]\,\,.
\label{ochenta}
\eeq
It is not difficult to obtain the action of the operators
of eq. (\ref{ochenta}) on the states $\xi_{j,k}(a,\tau)$.
The result is:
\beq
W_l^{(r,s)}\,\xi_{j,k}(a,\tau)\,=\,(-1)^{2l}\,
\sum_{n\in\Lambda_l}\,(-1)^{2n(s-1)}\,
q^{4rsn^2\,+\,2(4j+1)nr}\,\,
\xi_{j+ns,k}(a,\tau)\,\,,
\label{ouno}
\eeq
where $q$ is the same as in eq. (\ref{dseis}). In order to
prove eq. (\ref{ouno}) one has to use the well-known
behaviour of the theta functions under shifts in their
arguments. Notice that, remarkably, the action of the
operators (\ref{ochenta}) does not take us out of the
Hilbert space spanned by the functions (\ref{stsiete}).

Two particular cases of (\ref{ouno}) will be of great
interest for our purposes. First of all, let us consider
the situation in which $r=0$ and $s=1$, \ie\
a Wilson line along the $\beta$ cycle of the
torus. Eq. (\ref{ouno}) particularized to this case gives:
\beq
W_l^{(0,1)}\,\xi_{j,k}(a,\tau)\,=\,(-1)^{2l}\,
\sum_{n\in\Lambda_l}\,\xi_{j+n,k}(a,\tau)\,\,.
\label{odos}
\eeq
If one takes  $j=0$ in (\ref{odos}), which corresponds to
acting with $W_l^{(0,1)}$ on the vacuum state, one easily
arrives at:
\beq
W_l^{(0,1)}\,\xi_{0,k}(a,\tau)\,=\,(-1)^{2l}\,
\xi_{l,k}(a,\tau)\,\,.
\label{otres}
\eeq
In order to prove eq. (\ref{otres}) from eq. (\ref{odos}),
one has to make use of the periodicity properties of the
characters (eq. (\ref{trece})). Eq. (\ref{otres}) suggests
the interpretation of $W_l^{(0,1)}$ as a creation operator
of the state $\xi_{l,k}(a,\tau)$. Notice, however, the 
$(-1)^{2l}$ sign appearing in the right-hand side of eq. 
(\ref{otres}). We can absorb this sign by defining new
operators in the form:
\beq
\Phi_l^{(r,s)}\,\equiv\,(-1)^{2l}\,W_l^{(r,s)}\,\,.
\label{ocuatro}
\eeq
The operators $\Phi_j^{(r,s)}$   
are the Verlinde operators \cite{Verlinde} 
for the torus Hilbert space. Indeed, as the result of the
action of  $\Phi_j^{(0,1)}$ on the vacuum state
$\xi_{0,k}$,  the state $\xi_{j,k}$ of isospin $j$ is
obtained. Moreover, it can be checked from (\ref{odos})
that the operators  $\Phi_j^{(0,1)}$ satisfy the 
${\rm osp}(1\vert 2)$ fusion rules of eq. (\ref{seis}). It
is interesting to point out that, on the contrary,  the
Wilson line operators $W_l^{(0,1)}$ do not satisfy these
fusion rules. Actually, the composition law which they obey
has signs. The redefinition of eq. (\ref{ocuatro})  
eliminates these signs and, as a consequence,  
the correct fusion rules are reproduced. It is important to
point out here the difference with the situation for
bosonic gauge groups, where the representation of the
Verlinde operators is given directly by the Wilson lines. 

Another interesting particular case of eq. (\ref{ouno}) is 
$r=1$, $s=0$, which corresponds to  Wilson lines for the
$\alpha$ cycle of the torus. It follows from  (\ref{ouno})
that, in this case,  the Wilson line operators act
diagonally on the states (\ref{stsiete}):
\beq
W_l^{(1,0)}\,\xi_{j,k}(a,\tau)\,=\,
\sum_{n\in\Lambda_l}\,(-1)^{2(l+n)}\,
q^{2(4j+1)\,n}\,\xi_{j,k}(a,\tau)\,\,.
\label{ocinco}
\eeq
Remarkably enough, one can show that the alternate sum
appearing in the right-hand side of eq. (\ref{ocinco}) can
be put in terms of ratios of the entries of the modular
matrix $S$:
\beq
W_l^{(1,0)}\,\xi_{j,k}(a,\tau)\,=\,
(-1)^{2l}\,{S_{lj}\over S_{0j}}\,\
\xi_{j,k}(a,\tau)\,\,.
\label{oseis}
\eeq
Notice from eq. (\ref{oseis}) that the action of 
$W_l^{(1,0)}$ on the vacuum state $\xi_{0,k}$ is equivalent
to a multiplication of the latter by the quantum dimension 
$SD_q[l]$(see eq. (\ref{vdos})). Another interesting
consequence of eq. (\ref{oseis}) is the fact that the
Verlinde operators $\Phi_l^{(1,0)}$ also act diagonally on
the states (\ref{stsiete}). This is, actually, the content
of the Verlinde theorem, which states that the modular $S$
matrix diagonalizes the fusion rules. 

Our formalism can be used to compute vacuum expectation
values of Wilson lines on the three-sphere $S^3$. In this
calculation, we shall make use of the well-known fact that
the three-sphere $S^3$ can be obtained by joining together
two solid tori whose boundaries are identified by means of
a modular $S$ transformation. This $S$ transformation has
a well-defined realization in our Hilbert space.
Actually, its matrix elements are precisely the ones
displayed in eq. (\ref{quince}). The expectation values of
Wilson line operators in the vacuum, when the total three
manifold is $S^3$, can be simply obtained  by inserting the
$S$ transformation as follows:
\beq
<\,W_j^{(r,s)}\,>_{S^3}\,\,=\,\,
{(\,SW_j^{(r,s)}\,)_{00}\over S_{00}}\,\,.
\label{osiete}
\eeq
Notice that, in eq. (\ref{osiete}), we have normalized the
vacuum expectation values in such a way that they take the
value one for the unit operator (\ie\ when $j=0$ in 
(\ref{osiete})). It is understood in the right-hand side of
eq. (\ref{osiete}) that one is taking the diagonal matrix
element with respect to the vacuum state $\xi_{0,k}$. Using
the results of eqs. (\ref{quince}) and (\ref{ouno}), it is
straightforward to compute these matrix elements. One gets:
\beq
<\,W_j^{(r,s)}\,>_{S^3}\,\,=\,\,
\sum_{n\in\Lambda_j}\,(-1)^{2(j+n)}\,
q^{4rsn^2+2nr}\,\,
{\,q^{{4sn+1\over 2}}\,+\,q^{-{4sn+1\over 2}}\over
q^{{1\over 2}}\,+\,q^{-{1\over 2}}}\,\,.
\label{oocho}
\eeq
The sum over $\Lambda_j$ appearing in the right-hand side
of eq. (\ref{oocho}) can be done explicitly in some cases.
For example, if $s=1$, \ie\ for torus knots of type $(r,1)$,
one can easily verify that eq. (\ref{oocho}) reduces to:
\beq
<\,W_j^{(r,1)}\,>_{S^3}\,\,=\,\,q^{2j(2j+1)r}\,
{\,q^{{4j+1\over 2}}\,+\,q^{-{4j+1\over 2}}\over
q^{{1\over 2}}\,+\,q^{-{1\over 2}}}\,\,.
\label{onueve}
\eeq
Eq. (\ref{onueve}) contains very interesting information.
Let us take, first of all, the case $r=0$. The $(0,1)$
torus knot is nothing but the unknot. On the other hand, it
is evident from (\ref{onueve}) and (\ref{dsiete}) that:
\beq
<\,W_j^{{\rm unknot}}\,>_{S^3}\,=\,
\Bigl[\,{4j+1\over2}\,\Bigr]_+
\,=\,SD_q[j]\,\,.
\label{noventa}
\eeq
Therefore, as  happens for the CS theories with bosonic
gauge groups, the expectation values of unknot Wilson lines
are the quantum dimensions. Notice, however, that these
expectation values are not given by ratios of  $S$
matrix elements (see eq. (\ref{vdos})). This only occurs
when we take expectation values of the Verlinde operators 
(\ref{ocuatro}) for the unknot. It is also interesting to
look at the $r$ dependence of the right-hand side of eq. 
(\ref{onueve}). This dependence can be written as:
\beq
<\,W_j^{(r,1)}\,>_{S^3}\,\,=\,\,
e^{2\pi ih_j r}\,<\,W_j^{(0,1)}\,>_{S^3}\,\,,
\label{nuno}
\eeq
where $h_j$ are the conformal weights (\ref{cinco}).

Two knots or links are isotopically equivalent if they can
be transformed into each other by means of a series of
moves. The notion of isotopic equivalence depends on
the type of moves considered as basic  
deformations. In knot theory, Reidemeister introduced three
basic moves (denoted usually by I, II and III) which define
an equivalence relation called ambient isotopy \cite{knot}. 
If one does not include the type I Reidemeister move in the
equivalence relation, another notion of topological
equivalence, the so-called regular isotopy, is 
defined \cite{knot}. The $(r,1)$ torus knot is ambient
isotopy equivalent to the unknot. This fact is illustrated
in figure 6, where it is shown how one can convert the 
$(r,1)$ torus knot into the unknot by means of  $r$ type I
Reidemeister moves. 

\begin{figure}
\centerline{\hskip-.8in \epsffile{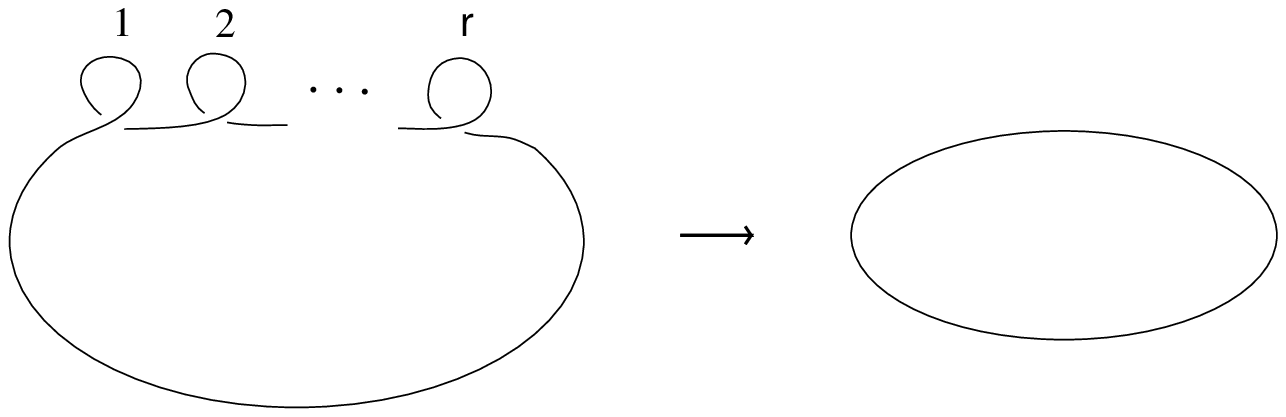}}
\caption{The $(r,1)$ torus knot is transformed into the
unknot by means of $r$  Reidemeister moves of type I}
\label{fig6}
\end{figure}

As is shown in eq. (\ref{nuno}), the vacuum expectation
value on $S^3$ of the Wilson line 
$W_j^{(r,1)}$ depends on $r$ and, therefore, this means that
$<\,W_j^{(r,1)}\,>_{S^3}$ is not invariant under ambient
isotopy. This is, actually, a general feature of CS gauge
theories which is usually interpreted as due to the fact
that the CS theory introduces a frame to the knots along
which Wilson lines are defined. Notice that the framing
dependence is a multiplicative factor depending on the
conformal weight $h_j$ of the ${\rm osp}(1\vert 2)$ current
algebra. This means that the effect of framing is
controlled by the monodromy behaviour of the CFT conformal
blocks. For a general $(r,s)$ torus knot, it is clear that
the effect due to framing will be a factor 
$exp[2\pi i rsh_j]=q^{2j(2j+1)rs}$. The knot polynomials
associated to our ${\rm osp}(1\vert 2)$ gauge theory can be
obtained by extracting this factor. They are polynomials in
the variable $q$ defined as:
\beq
P_j^{{(r,s)}}\,(q)\,\equiv\,q^{2j(2j+1)rs}\,\,
{<\,W_j^{(-r,s)}\,>_{S^3}\over
<\,W_j^{(0,1)}\,>_{S^3}}\,\,.
\label{ndos}
\eeq
Notice that in the right-hand side of eq. (\ref{ndos}) we
have changed the sign of $r$ in order to adequate our
orientation conventions  for torus knots to the standard
ones in the mathematics literature \cite{knot}. We have
normalized the knot polynomial in such a way that the
polynomial of the unknot is 1. Using the result written in
eq. (\ref{oocho}), it is now straightforward to obtain the
following expression of the polynomial for a general torus
knot:
\beq
P_j^{{(r,s)}}\,(q)\,=\,
{q^{2j(r-1)(s-1)}\over q^{4j+1}+1}\,\,
\sum_{p=0}^{4j}\,\,
(-1)^p\,q^{r(1+sp)(4j-p)}\,\,
(\,q^{1+sp}\,+\,q^{s(4j-p)}\,)\,\,.
\label{ntres}
\eeq
From eq. (\ref{ntres}), it is easy to find the relation 
between the ${\rm osp}(1\vert 2)$ and su$(2)$ polynomials.
It is interesting to point out that this relation can be
obtained by using the same identification of the
deformation parameters of $U_q({\rm osp}(1\vert 2))$ and
$U_t({\rm su}(2))$ that was found in sections 2 and 3.
Indeed, let  $\tilde P_{j}^{{(r,s)}}\,(t)$ be  the 
su$(2)$ polynomial of isospin $j$, in the variable $t$, for
an $(r,s)$ torus knot. The explicit expressions of the 
su$(2)$ polynomials for torus knots were obtained in ref.
\cite{poly}. Comparing these results with eq. (\ref{ntres}),
one easily realizes that:
\beq
P_j^{{(r,s)}}\,(q)\,=\,\tilde P_{2j}^{{(r,s)}}\,(-q)\,\,.
\label{ncuatro}
\eeq
Notice that eq. (\ref{ncuatro}) implies that our 
${\rm osp}(1\vert 2)$ polynomials can be identified with
the su$(2)$ polynomials for integer isospins. Therefore, for
example, the ${\rm osp}(1\vert 2)$ polynomial for the
fundamental representation is identified in eq. 
(\ref{ncuatro}) with the Akutsu-Wadati polynomial obtained
from the three-state vertex model (\ie\ from the
nineteen-vertex model) \cite{AW}.
So far we have only related ${\rm osp}(1\vert 2)$ and
su$(2)$  polynomials for torus knots. In next section we
shall be able to verify this relation for more general
classes of knots of links. Within our present formalism, we
can generalize eq. (\ref{ncuatro}) to arbitrary torus links.
Let us remember that, in general, an $(r,s)$ torus link can
be represented as the closure of the braid with $s$ strands 
$(\sigma_1\,\cdots\,\sigma_{s-1})^r$, where $\sigma_i$ is
the operation that interchanges the strands numbered $i$
and $i+1$. When $r$ and $s$ are coprime, the link is an 
$(r,s)$ torus knot. On the contrary, when the greatest
common divisor of $r$ and $s$, which we shall denote by 
$gcd(r,s)$, is different from one, the link has more that
one component. In fact, it is not difficult to convince
oneself that the number of  components of the $(r,s)$
torus link is given precisely by:
\beq
\nu_{r,s}\,=\,gcd(r,s)\,\,.
\label{ncinco}
\eeq
Furthermore, one can verify that each of the  $\nu_{r,s}$
components of the link is an 
$(r/\nu_{r,s}\,,\,s/\nu_{r,s})$ torus knot.

The polynomial for a link can be obtained by means of a
slight generalization of our prescription for knots 
(eq. (\ref{ndos})). Basically, one must compute the
expectation value of a product of more that one Wilson line
operators. Actually, one must insert in the correlator a
Wilson line operator for every component of the link. Taking
into account the framing factor, these considerations lead
us to define the polynomial for an $(r,s)$ 
torus link as \cite{poly}:
\beq
P^{(r,s)}_j(q)={q^{2j(2j+1)rs}\over 
{\langle W^{(0,1)}_{j}  \rangle }_{S^3}}\,\,
\Bigl \langle \,\Biggl (
W^{(-{r\over \nu_{r,s}}\,,\,{s\over
\nu_{r,s}})}_j \Biggr )^{\nu_{r,s}}\,
\Bigr \rangle _{S^3}.
\label{nseis}
\eeq
The calculation of the right-hand side of eq. (\ref{nseis})
can be performed by using the same techniques as in ref.
\cite{poly}. If $\tilde P^{(r,s)}_{j}(t)$ is the su$(2)$
polynomial for an $(r,s)$ torus knot in the isospin $j$
representation, one can prove that:
\beq
P^{(r,s)}_j(q)\,=\,(-1)^{2j(\nu_{r,s}-1)}\,\,
\tilde P^{(r,s)}_{2j}(-q)\,\,.
\label{nsiete}
\eeq
The values of the su$(2)$ polynomials for torus links were
given in ref. \cite{poly} and will not be reproduced here. 
Notice that now, in eq. (\ref{nsiete}), apart from the 
$q\rightarrow -q$, $j\rightarrow 2j$ correspondence, there
appears a $(-1)^{2j}$ sign, which was not present in the
case of knots. In next section we shall develop an
approach which will allow us to confirm this 
${\rm osp}(1\vert 2)/{\rm su}(2)$ connection for arbitrary
links.

\setcounter{equation}{0}
\section{osp$\bf{(1\vert 2)}$ invariants for arbitrary
knots and links}

The topological nature of CS gauge theories allows to
describe a given three-dimensional situation in terms of
different two-dimensional problems. Exploiting this
richness of the CS theories, a series of
powerful computational techniques can be developed. Indeed,
as we recalled at the beginning of section 4, in the
quantization of the CS theory on the three-sphere, one must
split 
$S^3$ as a connected sum of two
three-manifolds with a common boundary $\Sigma$. In section
4 we have been considering the particular case in which 
$\Sigma=T^2$ and there are no Wilson lines cut by the
intermediate torus. In the present section,  we shall cut
the 
$S^3$ manifold along a two-sphere $S^2$,  chosen in such a
way that it intersects the Wilson lines in four points. In
order to characterize the CS states in these punctured
two-spheres,  we shall make use of the results of section 3
on the behaviour under crossing symmetry of the four-point
conformal blocks in  the ${\rm osp}(1\vert 2)$ CFT. 

\begin{figure}
\centerline{\hskip.4in \epsffile{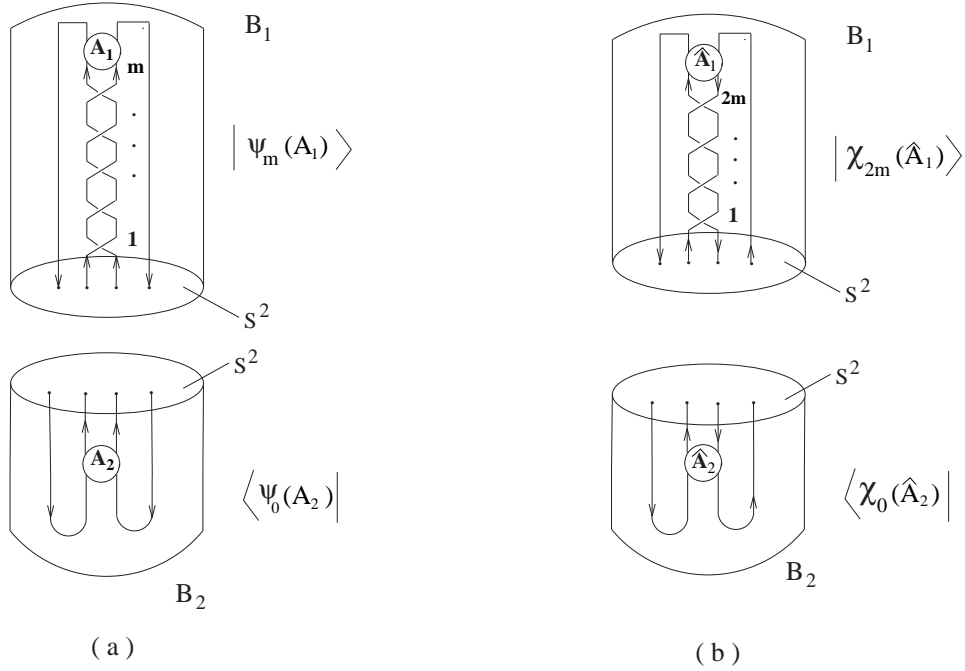}}
\caption{Composition of balls $B_1$ and $B_2$ which yields
the links $L_m\,(A_1\,,\,A_2\,)$ (a) and
$\hat L_{2m}\,(\hat A_1\,,\,\hat A_2\,)$ (b) in $S^3$.}
\label{fig7}
\end{figure}

Let us build up our formalism, following ref. \cite{kaul}.
First of all,  we shall introduce some definitions. We shall
call  a compact three-dimensional submanifold in
$S^3$ with some points of the boundary marked as ``in" or
``out",  a ``room" \cite{millet}. An ``inhabitant" of the
room is, by definition, a properly embedded smooth, compact
oriented one-dimensional manifold which meets the boundary
of the room at the given set of marked points with its
orientation matching the  ``in" and  ``out" designations.
Given two rooms $A_1$ and
$A_2$ with two ``in" and two ``out" points, let us consider
the link $L_m\,(A_1\,,\,A_2\,)$, obtained by joining $A_1$ 
and $A_2$ by four strands, with $m$ half-twists in two of
the parallelly oriented strands. In figure \ref{fig7}a,  we
have represented the embedding of $L_m\,(A_1\,,\,A_2\,)$ in 
$S^3$. As shown in this figure,  we shall decompose $S^3$
into two solid balls $B_1$ and $B_2$ in such a way that
$B_1$ contains the room $A_1$ and the $m$ half-twists in
its parallel strands and $B_2$ contains the room $A_2$.
Notice that the common boundary $S^2$ intersects with the
four strands and that the lower two strands of $A_1$ and
$A_2$ are parallelly oriented. We shall also consider the
links $\hat L_{2m}\,(\hat A_1\,,\,\hat A_2\,)$, where 
$\hat A_1$ and $\hat A_2$ are rooms whose lower two strands
have opposite orientation. The link 
$\hat L_{2m}\,(\hat A_1\,,\,\hat A_2\,)$ is obtained, as
shown in figure \ref{fig7}b, by joining the rooms 
$\hat A_1$ and $\hat A_2$ with four strands, with $2m$
half-twists in the oppositely oriented lower two strands of 
$\hat A_1$.

Let us now describe how one can obtain the 
${\rm osp}(1\vert 2)$ polynomials for the links of figure 
\ref{fig7}. We are going to consider first the link 
$L_m\,(A_1\,,\,A_2\,)$. As shown in figure \ref{fig7}a, one
can associate a state to each of the two balls $B_1$ and
$B_2$ in which $S^3$ is split. These states, denoted by 
$|\,\psi_m(A_1)\,>$ and $<\psi_0(A_2)\,|$, can be obtained
by performing the CS functional integral over the balls 
$B_1$ and $B_2$ respectively. The expectation value of the
link will be given by the inner product of these two states.
The corresponding invariant polynomial can be obtained after
extracting the framing dependence and after performing a
convenient normalization. Therefore, if we denote the isospin
$j$ polynomial of the link by
$P_j\,[\,L_m\,(A_1\,,\,A_2\,)\,]\,(q)$, it is clear that:
\beq
P_j\,[\,L_m\,(A_1\,,\,A_2\,)\,]\,(q)\,=\,
{<\psi_0(A_2)\,|\,\psi_m(A_1)\,>\over SD_q[j]}\,\,.
\label{nocho}
\eeq
In eq. (\ref{nocho}),  we have taken into account the result
(\ref{noventa}) for the expectation value of the unknot. The
dependence of the right-hand side of eq. (\ref{nocho}) 
in the number $m$ of twists can
be obtained as follows. Let $B_j^{(+)}$ be the operator
that introduces a frame corrected half twist in the middle
two strands of the CS states on $S^2$.
Obviously, the action of $B_j^{(+)}$ on the states is:
\beq
B_j^{(+)}\,|\,\psi_l(A_1)\,>\,=\,|\,\psi_{l+1}(A_1)\,>\,\,.
\label{nnueve}
\eeq
It is interesting for our purposes to introduce the
eigenvectors $|\,\phi_l^{(+)}\,>$ of the $B_j^{(+)}$
operator:
\beq
B_j^{(+)}\,|\,\phi_l^{(+)}\,>\,=\,
\Lambda_{l,j}^{(+)}\,(q)\,|\,\phi_l^{(+)}\,>\,\,.
\label{cien}
\eeq
From our analysis of the four-point conformal blocks of the 
${\rm osp}(1\vert 2)$ CFT (sect. 3), it is clear that
$B_j^{(+)}$ acts in a $4j+1$-dimensional space. Actually,
we shall label the braiding eigenstates and eigenvalues as
in section 3 and, therefore, $l$ in eq. (\ref{cien}) will
take the values $l\,=\,0,\cdots,2j$ with $2l\in\ZZ$.  Let
us denote the braiding eigenvalues of eq. (\ref{ciuno}) for 
$j_1=j_2=j$ and $j_p=l$ by $\Lambda_j^l(q)$. After taking
into account the framing corrections (eq. (\ref{nuno})) and
our orientation conventions (eq. (\ref{ndos})), it is clear
that:
\beq
\Lambda_{l,j}^{(+)}(q)\,=\,exp[\,2\pi i h_j\,]\,
\Lambda_j^l(q^{-1})\,\,.
\label{ctuno}
\eeq
Therefore, making use of eq. (\ref{ciuno}), one can write:
\bear
\Lambda_{l,j}^{(+)}(q)&=&
(-1)^{<2j-l+{1\over 2}>}\,q^{4j(2j+1)-l(2l+1)}\rc\rc
l\,&=&\,0,\cdots,2j,\,\,\,\,\,\,\,\,\,\,\,\,\,\,\,\,\,\,\,
2l\in\ZZ\,\,.\label{ctdos}\\
\nonumber
\eear
With the eigenvalues (\ref{ctdos}) at our disposal, we can
use the characteristic equation of $B_j^{(+)}$ in order to
get recursion relations (skein rules) that the 
${\rm osp}(1\vert 2)$ polynomials must obey. However, these
skein rules, which relate the polynomials (\ref{nocho}) for
different values of $m$, cannot completely determine the CS
invariants for arbitrary knots and links. Fortunately, as
was shown in ref. \cite{kaul}, the polynomials can be
directly obtained from the monodromy properties of the
two-dimensional four-point correlators. Indeed, if we
expand the vectors appearing in the decomposition of figure 
\ref{fig7}a for $m=0$ as:
\bear
|\,\psi_0(A_1)\,>\,&=&\,\sum_{{l=0\atop 2l\in\ZZ}}^{2j}\,\,
\mu_l^{(+)}(A_1)\,|\,\phi_l^{(+)}\,>\rc\rc
<\,\psi_0(A_2)\,|\,&=&\,\sum_{{l=0\atop 2l\in\ZZ}}^{2j}\,\,
\mu_l^{(+)}(A_2)\,<\,\phi_l^{(+)}\,|\,\,,
\label{cttres}\\
\nonumber
\eear
then, using eqs. (\ref{nnueve}) and (\ref{cien}), the state 
$|\,\psi_m(A_1)\,>$ can be written as:
\beq
|\,\psi_m(A_1)\,>\,=\,\sum_{{l=0\atop 2l\in\ZZ}}^{2j}\,\,
\mu_l^{(+)}(A_1)\,[\,\Lambda_{l,j}^{(+)}\,(q)\,]^m\,
|\,\phi_l^{(+)}\,>\,\,.
\label{ctcuatro}
\eeq
In eq. (\ref{cttres}),  $\mu_l^{(+)}(A_1)$ and 
$\mu_l^{(+)}(A_2)$ are certain coefficients which depend on
the rooms $A_1$ and $A_2$. The states
$|\,\phi_l^{(+)}\,>$, $l=0,\cdots, 2j$ can be chosen in
such a way that they form a complete orthonormal set.
Therefore, their inner product with the elements 
$<\,\phi_l^{(+)}\,|$ of the dual basis is given by 
$<\,\phi_m^{(+)}\,|\,\phi_l^{(+)}\,>\,=\,\delta_{ml}$.
Using this fact, one can write the following expression for
the polynomial of the link $L_m\,(A_1\,,\,A_2\,)$:
\beq
P_j\,[\,L_m\,(A_1\,,\,A_2\,)\,]\,(q)\,=\,
{1\over SD_q[j]}\,\,
\sum_{{l=0\atop 2l\in\ZZ}}^{2j}\,\,
\mu_l^{(+)}(A_1)\,\mu_l^{(+)}(A_2)\,
[\,\Lambda_{l,j}^{(+)}\,(q)\,]^m\,\,.
\label{ctcinco}
\eeq
Notice that, in eq. (\ref{ctcinco}), the dependence of 
$P_j\,[\,L_m\,(A_1\,,\,A_2\,)\,]\,(q)$ on the number $m$ of
half-twists has been explicitly determined. However, we
need to obtain the coefficients appearing in the linear
combinations (\ref{cttres}) in order to get the
actual expression of the polynomial. Let us consider, first
of all, the case in which $A_1$ and $A_2$ are the
``trivial" rooms, \ie\ when  $A_1$ and $A_2$ are:
\beq
\centerline{\hskip.4in \epsffile{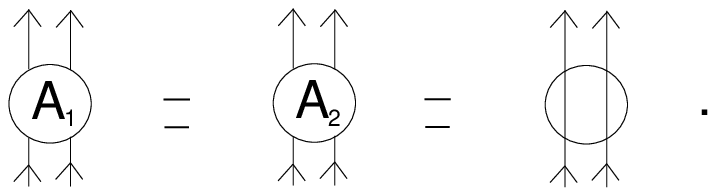}}
\label{ctseis}
\eeq
When $A_1$ and $A_2$ are the rooms (\ref{ctseis}), the link 
$L_m\,(A_1\,,\,A_2\,)$ is simply the link 
${\cal L}_m$ obtained as the
closure of an $m$-twisted braid of two parallelly oriented
strands.  Notice that ${\cal L}_m$ is nothing but the
$(m,2)$ torus link, which has one(two) components when $m$
is odd(even). If we denote by $\mu_l^{(+)}$ the
coefficients $\mu_l^{(+)}(A_1)$ and 
$\mu_l^{(+)}(A_2)$ when $A_1$ and $A_2$ are given by
(\ref{ctseis}), it is evident that, in this case, eq. 
(\ref{ctcinco}) reduces to:
\beq
P_j^{(m,2)}(q)\,=\,P_j\,[\,{\cal L}_m\,](q)\,=\,
{1\over SD_q[j]}\,\,
\sum_{l=0\atop 2l\,\in\,\ZZ}^{2j}\,
[\,\mu_l^{(+)}\,]^2\,[\,\Lambda_{l,j}^{(+)}(q)\,]^{m}\,\,.
\label{ctsiete}
\eeq
We have determined the expression of $P_j^{(m,2)}(q)$ in
section 4 (see eq. (\ref{ntres}) for knots and eq.
(\ref{nsiete}) in the case of  links). It can be easily
proved that the results of section 4 can be written as:
\beq
P_j^{(m,2)}(q)\,=\,
{1\over SD_q[j]}\,\,
\sum_{l=0\atop 2l\,\in\,\ZZ}^{2j}\,
(-1)^{2l}\,\Bigl[\,{4l+1\over 2}\,\Bigr]_+\,
(-1)^{m<2j-l+{1\over 2}>}\,
q^{m[4j(2j+1)-l(2l+1)]}\,\,.
\label{ctocho}
\eeq
In the right-hand side of eq. (\ref{ctocho}), one can
immediately recognize the $m^{{\rm th}}$ power of the
braiding eigenvalue $\Lambda_{l,j}^{(+)}(q)$. Therefore, the 
$\mu_l^{(+)}$ coefficients satisfy:
\beq
[\,\mu_l^{(+)}\,]^2\,=\,(-1)^{2l}\,
\Bigl[\,{4l+1\over 2}\,\Bigr]_+\,\,,
\label{ctnueve}
\eeq
which means that the $\mu_l^{(+)}$ are given by:
\beq
\mu_l^{(+)}\,=\,i^{2(l-<l>)}\,
\sqrt{\Bigl[\,{4l+1\over 2}\,\Bigr]_+}\,\,.
\label{ctdiez}
\eeq
Later in this section we shall use the result
(\ref{ctdiez}) to determine the CS states associated to
general classes of rooms.

Let us now consider the link 
$\hat L_m\,(\hat A_1\,,\,\hat A_2\,)$ represented in 
\ref{fig7}b. It is clear that, in this case, the
corresponding polynomial is:
\beq
P_j\,[\,\hat L_{2m}\,(\hat A_1\,,\,\hat A_2\,)\,]\,(q)\,=\,
{<\chi_0(\hat A_2)\,|\,\chi_{2m}(\hat A_1)\,>\over SD_q[j]}
\,\,.
\label{ctonce}
\eeq
We can evaluate the right-hand side of eq. (\ref{ctonce}) 
following the same strategy used to arrive at eq. 
(\ref{ctcinco}). Let us introduce the operator 
$B_j^{(-)}$ that implements the half-twists in the 
oppositely oriented middle two strands on the ball $B_1$ in
figure \ref{fig7}b. As in the case of $B_j^{(+)}$, we shall
assume that in the action of $B_j^{(-)}$ the framing
dependence has been eliminated. In analogy with eq. 
(\ref{nnueve}), one has 
$|\,\chi_{2m}(\hat A_1)\,>\,=\,[\,B_j^{(-)}\,]^{2m}\,
|\,\chi_{0}(\hat A_1)\,>$. Let us also introduce a
complete set of orthonormal eigenstates of $B_j^{(-)}$:
\beq
B_j^{(-)}\,|\,\phi_l^{(-)}\,>\,=\,
\Lambda_{l,j}^{(-)}\,(q)\,|\,\phi_l^{(-)}\,>\,\,.
\label{ctdoce}
\eeq
In terms of the $|\,\phi_l^{(-)}\,>$'s, the zero-twist
states $|\,\chi_{0}(\hat A_1)\,>$ and 
$<\,\chi_{0}(\hat A_2)\,|$ can be expanded as follows:
\bear
|\,\chi_0(\hat A_1)\,>\,&=&\,\sum_{{l=0\atop
2l\in\ZZ}}^{2j}\,\,
\mu_l^{(-)}(\hat A_1)\,|\,\phi_l^{(-)}\,>\rc\rc
<\,\chi_0(\hat A_2)\,|\,&=&\,\sum_{{l=0\atop
2l\in\ZZ}}^{2j}\,\,
\mu_l^{(-)}(\hat A_2)\,<\,\phi_l^{(-)}\,|\,\,,
\label{cttrece}\\
\nonumber 
\eear
and the polynomial 
$P_j\,[\,\hat L_{2m}\,(\hat A_1\,,\,\hat A_2\,)\,]\,(q)$ is
given by:
\beq
P_j\,[\,\hat L_{2m}\,(\hat A_1\,,\,\hat A_2\,)\,]\,(q)\,=\,
{1\over SD_q[j]}\,\,
\sum_{{l=0\atop 2l\in\ZZ}}^{2j}\,\,
\mu_l^{(-)}(\hat A_1)\,\mu_l^{(-)}(\hat A_2)\,
[\,\Lambda_{l,j}^{(-)}\,(q)\,]^{2m}\,\,.
\label{ctcatorce}
\eeq
The eigenvalues $\Lambda_{l,j}^{(-)}\,(q)$, which determine
the $m$ dependence in the right-hand side of eq. 
(\ref{ctcatorce}),  can be determined 
as was done in ref. \cite{kaul} 
for the su$(2)$ case. Actually, as now one of the
directions of the two strands is reversed, it is easy to
convince oneself that the $\Lambda_{l,j}^{(-)}\,(q)$'s are
given by:
\beq
\Lambda_{l,j}^{(-)}(q)\,=\,(-1)^{2j}\,exp[\,2\pi i h_j\,]\,
\Lambda_j^l(q)\,\,.
\label{ctquince}
\eeq
Using eq. (\ref{ciuno}) and the relation 
$(-1)^{2j} (-1)^{<2j-l+{1\over 2}>}\,=\,(-1)^{<l>}$, it is 
straightforward to arrive at the following expression:
\bear
\Lambda_{l,j}^{(-)}(q)\,&=&\,(-1)^{<l>} q^{l(2l+1)}\rc\rc
l\,&=&\,0,\cdots,2j\,\,\,\,\,\,\,\,\,\,\,\,\,\,
2l\in\ZZ\,\,.
\label{ctdseis}\\
\nonumber
\eear
Let us now restrict ourselves to the case in which the
rooms $\hat A_1$ and $\hat A_2$ are given by:
\beq
\centerline{\hskip.4in \epsffile{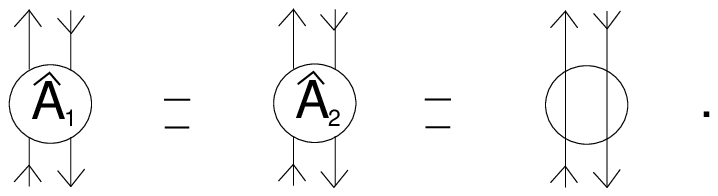}}
\label{ctdsiete}
\eeq
In this case, $\hat L_{2m}\,(\hat A_1\,,\,\hat A_2\,)$ is
simply the link $\hat {\cal L}_{2m}$ obtained as the closure
of two oppositely oriented strands with $2m$ twists. If we
denote by $\mu_l^{(-)}$ the coefficients 
$\mu_l^{(-)}(\hat A_1)$ and $\mu_l^{(-)}(\hat A_2)$ when 
$\hat A_1$ and $\hat A_2$ are the rooms 
(\ref{ctdsiete}), it follows from eqs. (\ref{ctcatorce})
and (\ref{ctdseis}) that the polynomial of the link 
$\hat {\cal L}_{2m}$ can be written as:
\bear
P_j\,[\,\hat {\cal L}_{2m}\,](q)\,=\,
{1\over SD_q[j]}\,\,
\sum_{l=0\atop 2l\,\in\,\ZZ}^{2j}\,
[\,\mu_l^{(-)}\,]^2\,[\,\Lambda_{l,j}^{(-)}(q)\,]^{2m}\,=\,
{1\over SD_q[j]}\,\,
\sum_{l=0\atop 2l\,\in\,\ZZ}^{2j}\,
[\,\mu_l^{(-)}\,]^2\,q^{2ml(2l+1)}\,\,.\rc
\label{ctdocho}
\eear
The link $\hat {\cal L}_{+2}$ ($\hat {\cal L}_{-2}$) is
nothing but the right(left)-handed  Hopf link $H$ ($H^*$).
The expectation value for the Wilson lines corresponding to
$H$ and $H^*$ can be obtained from our results of section 4.
Indeed, it follows from the expression of the Verlinde
operators in the torus Hilbert space (eq. (\ref{ocuatro}))
that the expectation value for these links is 
$S_{jj}/S_{00}$, where the $S_{ij}$'s are given by eq. 
(\ref{quince}). After taking into account the framing
corrections and our normalization conventions for the
invariant polynomial, we can write:
\beq
P_j[H](q)\,=\,{e^{4\pi i h_j}\over SD_q[j]}\,\,
{S_{jj}\over S_{00}}\,=\,
{q^{4j(2j+1)}\over SD_q[j]}\,
\,\Bigl[\,{(4j+1)^2\over 2}\,\Bigr]_+\,\,.
\label{ctdnueve}
\eeq
After some manipulations of the graded quantum numbers,
eq. (\ref{ctdnueve}) can be recast as:
\beq
P_j[H](q)\,=\,{1\over SD_q[j]}\,\,
\sum_{l=0\atop 2l\,\in\,\ZZ}^{4j(2j+1)}\,
(-1)^{2l}\,q^{2l}\,=\,{1\over SD_q[j]}\,\,
\sum_{l=0\atop 2l\,\in\,\ZZ}^{2j}\,
(-1)^{2l}\,\Bigl[\,{4l+1\over 2}\,\Bigr]_+\,\,
q^{2l(2l+1)}\,\,.
\label{ctveinte}
\eeq
The polynomial for the left-handed Hopf link can be obtained
by changing  $q\rightarrow q^{-1}$, namely:
\beq
P_j[H^*](q)\,=\,{1\over SD_q[j]}\,\,
\sum_{l=0\atop 2l\,\in\,\ZZ}^{2j}\,
(-1)^{2l}\,\Bigl[\,{4l+1\over 2}\,\Bigr]_+\,\,
q^{-2l(2l+1)}\,\,.
\label{ctvuno}
\eeq
Eqs. (\ref{ctveinte}) and (\ref{ctvuno}) should correspond
to eq. (\ref{ctdocho}) when $m=\pm1$. By a simple
inspection of these equations one can determine the values
of the $ \mu_l^{(-)}$ coefficients. The result that one
gets is:
\beq
 \mu_l^{(-)}\,=\,i^{2(l-<l>)}\,
\sqrt{\Bigl[\,{4l+1\over 2}\,\Bigr]_+}\,\,.
\label{ctvdos}
\eeq
Therefore, substituting eq. (\ref{ctvdos}) in eq. 
(\ref{ctdocho}), the following expression for 
$P_j[\,\hat{\cal L}_{2m}\,](q)$ is obtained:
\beq
P_j[\,\hat{\cal L}_{2m}\,](q)\,=\,{1\over SD_q[j]}\,\,
\sum_{l=0\atop 2l\,\in\,\ZZ}^{2j}\,
(-1)^{2l}\,
\Bigl[\,{4l+1\over 2}\,\Bigr]_+\,
\,q^{2ml(2l+1)}\,\,.
\label{ctvtres}
\eeq

\begin{figure}
\centerline{\hskip.4in \epsffile{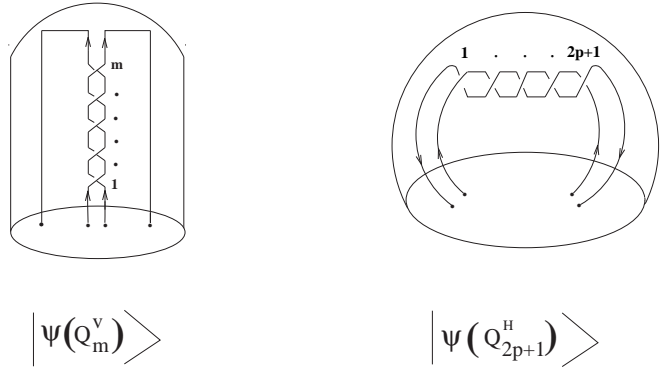}}
\caption{Balls containing the rooms $Q_m^V$ 
and $Q_{2p+1}^H$.}
\label{fig10}
\end{figure}

The results found so far in this section can be used to
determine the CS states corresponding to the  rooms
$Q_m^V$ and $Q_{2p+1}^H$, displayed in figure 
\ref{fig10}. In general, these states can be given as a
linear combination of the elements of the basis 
$\{|\,\phi_l^{(+)}\,>\}$:
\bear
|\,\psi\,(Q^{V}_{m})\,>\,&=&\,
\sum_{l=0\atop 2l\,\in\,\ZZ}^{2j}\,
\mu_l(Q^{V}_{m})\,|\,\phi_l^{(+)}\,>\rc\rc
|\,\psi\,(Q^{H}_{2p+1})\,>\,&=&\,
\sum_{l=0\atop 2l\,\in\,\ZZ}^{2j}\,
\mu_l(Q^{H}_{2p+1})\,|\,\phi_l^{(+)}\,>\,\,.
\label{ctvcuatro}\\
\nonumber
\eear
Let us now explain how the coefficients in
(\ref{ctvcuatro}) can be determined. We are going to
consider first the case of $Q_m^V$. It is clear that 
$|\,\psi\,(Q^{V}_{m})\,>$ should coincide with 
$|\,\psi_m\,(A_{1})\,>$ when $A_1$ is the room of eq. 
(\ref{ctseis}). This observation immediately implies that:
\beq
|\,\psi\,(Q^{V}_{m})\,>\,=\,
\sum_{l=0\atop 2l\,\in\,\ZZ}^{2j}\,
\mu_l^{(+)}\,[\,\Lambda_{l,j}^{(+)}(q)\,]^m
\,|\,\phi_l^{(+)}\,>\,\,.
\label{ctvcinco}
\eeq
and, therefore, $\mu_l(Q^{V}_{m})$ can be written as:
\beq
\mu_l(Q^{V}_{m})\,=\,i^{2(l-<l>)}\,
\sqrt{\Bigl[\,{4l+1\over 2}\,\Bigr]_+}\,\,\,
(\,\Lambda_{l,j}^{(+)}(q)\,)^{m}\,\,.
\label{ctvseis}
\eeq

The room $Q_{2p+1}^H$ has $2p+1$ half-twists in the first
two strands on the left (see figure \ref{fig10}). Therefore,
it is more convenient in this case to expand the state 
$|\,\psi\,(Q^{H}_{2p+1})\,>$ in terms of a basis in which
the action of the braiding of the first two strands on the
left is diagonal. Let us denote the elements of such a
basis by  $\,|\,\tilde \phi_l^{(-)}\,>$. Notice that the
two strands which are braided in  $Q_{2p+1}^H$  are
antiparallel and, thus, each half twist will introduce a
factor $\Lambda_{l,j}^{(-)}(q)$ multiplying the 
$l^{{\rm th}}$ eigenvector. Proceeding as before, it is easy
to obtain the vector $|\,\psi\,(Q^{H}_{2p+1})\,>$ in terms
of the $\,|\,\tilde \phi_l^{(-)}\,>$'s. One gets:
\beq
|\,\psi\,(Q^{H}_{2p+1})\,>\,=\,
\sum_{l=0\atop 2l\,\in\,\ZZ}^{2j}\,
\mu_l^{(-)}\,[\,\Lambda_{l,j}^{(-)}(q)\,]^{2p+1}
\,|\,\tilde \phi_l^{(-)}\,>\,\,.
\label{ctvsiete}
\eeq
The basis $\{|\,\tilde \phi_l^{(-)}\,>\}$, referring to
the first two strands on the left, and the one constituted
by the vectors $|\,\phi_l^{(+)}\,>$, which are eigenvectors
of the braiding operator of the middle two strands, must be
linearly related. Let us represent this relation as:
\beq
|\,\tilde \phi_r^{(-)}\,>\,=\,
\sum_{l=0\atop 2l\,\in\,\ZZ}^{2j}\,
a_{rl}\,|\,\phi_l^{(+)}\,>\,\,.
\label{ctvocho}
\eeq
The matrix 
$a_{rl}= <\,\phi_l^{(+)}|\,\tilde \phi_r^{(-)}\,>$ is a
duality matrix whose explicit expression shall be
determined below. Substituting eq. (\ref{ctvocho}) in eq. 
(\ref{ctvsiete}), one can get the value of the coefficients 
$\mu_l(Q^{H}_{2p+1})$ appearing in eq. (\ref{ctvcuatro}):
\beq
\mu_l(Q^{H}_{2p+1})\,=\,
\sum_{r=0\atop 2r\,\in\,\ZZ}^{2j}\,
i^{2(r-<r>)}\,
\sqrt{\Bigl[\,{4r+1\over 2}\,\Bigr]_+}\,\,\,
(\,\Lambda_{r,j}^{(-)}(q)\,)^{2p+1}\,
a_{rl}\,\,.
\label{ctvnueve}
\eeq
\begin{figure}
\vskip-.2in
\centerline{\epsffile{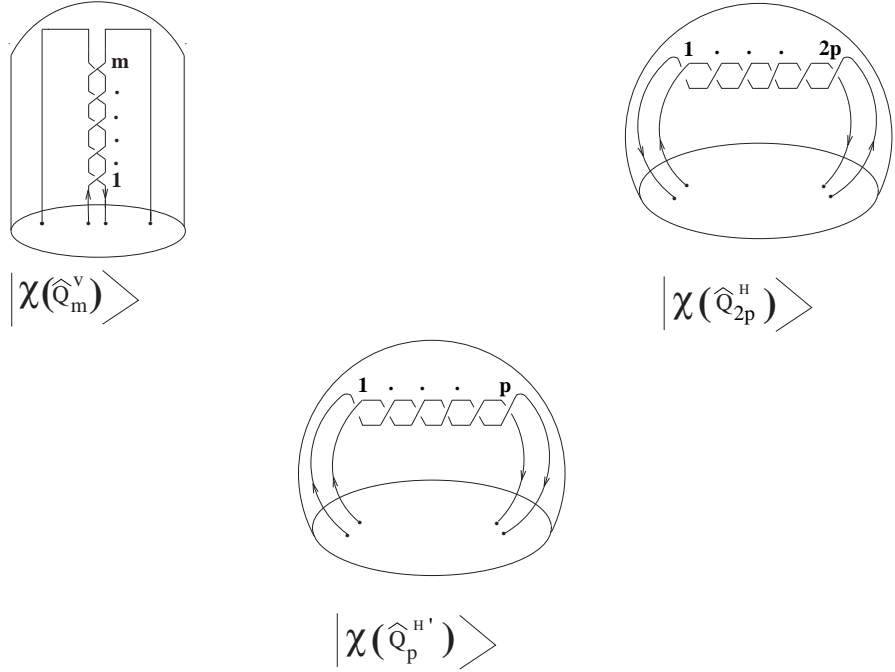}}
\caption{Balls containing the rooms 
$\hat Q^{\,V}_{m}$, $\hat Q^{\,H}_{2p}$ and 
$\hat Q^{\,H'}_{p}$.}
\label{fig11}
\end{figure}

Following the same methods, one can obtain the CS states
corresponding to the rooms $\hat Q^{\,V}_{m}$, 
$\hat Q^{\,H}_{2p}$ and $\hat Q^{\,H'}_{p}$ (see figure 
\ref{fig11}). Let us express them in terms of the basis 
 $\{|\,\phi_l^{(-)}\,>\}$, defined in eq. (\ref{ctdoce}):
\bear
|\,\chi\,(\hat Q^{\,V}_{m})\,>\,&=&\,
\sum_{l=0\atop 2l\,\in\,\ZZ}^{2j}\,
\hat \mu_l(\hat Q^{\,V}_{m})\,|\,
\phi_l^{(-)}\,>\rc 
|\,\chi\,(\hat Q^{\,H}_{2p})\,>\,
&=&\, \sum_{l=0\atop 2l\,\in\,\ZZ}^{2j}\, 
\hat \mu_l(\hat
Q^{\,H}_{2p})\,|\,\phi_l^{(-)}\,>\rc 
|\,\chi\,(\hat Q^{\,H'}_{p})\,>\,
&=&\, \sum_{l=0\atop 2l\,\in\,\ZZ}^{2j}\, 
\hat \mu_l(\hat
Q^{\,H'}_{p})\,|\, \phi_l^{(-)}\,>\,\,.
\label{cttreinta}\\
\nonumber
\eear

The coefficients $\hat \mu_l(\hat Q^{\,V}_{m})$, 
$\hat \mu_l(\hat Q^{\,H}_{2p})$ and 
$\hat \mu_l(\hat Q^{\,H'}_{p})$ entering the linear
combinations (\ref{cttreinta}) are given by:
\bear
\hat \mu_l(\hat Q^{\,V}_{m})\,&=&\,
i^{2(l-<l>)}\,
\sqrt{\Bigl[\,{4l+1\over 2}\,\Bigr]_+}\,\,\,
(\,\Lambda_{l,j}^{(-)}(q)\,)^{m}\rc
\hat \mu_l(\hat Q^{\,H}_{2p})\,&=&\,
\sum_{r=0\atop 2r\,\in\,\ZZ}^{2j}\,
i^{2(r-<r>)}\,
\sqrt{\Bigl[\,{4r+1\over 2}\,\Bigr]_+}\,\,\,
(\,\Lambda_{r,j}^{(-)}(q)\,)^{2p}\,\,
a_{rl}\rc
\hat \mu_l(\hat Q^{\,H'}_{p})\,&=&\,
\sum_{r=0\atop 2r\,\in\,\ZZ}^{2j}\,
i^{2(r-<r>)}\,
\sqrt{\Bigl[\,{4r+1\over 2}\,\Bigr]_+}\,\,\,
(\,\Lambda_{r,j}^{(+)}(q)\,)^{p}\,\,
a_{rl}\,\,.\label{cttuno}\\
\nonumber
\eear

The matrix $a_{rl}$ in eq. (\ref{cttuno}) is the same as in
eq. (\ref{ctvnueve}), \ie\ is the matrix that relates the
braiding in the first two strands to the braiding in the
middle two strands. It is clear from its definition that
$a_{rl}$ should be related to some of the duality matrices
we have found in section 3 for the ${\rm osp}(1\vert 2)$
CFT. There are several non-trivial requirements that 
$a_{rl}$ must satisfy. These requirements can be obtained
as consistency checks of our formalism. Indeed, taking the
rooms of figures \ref{fig10} and \ref{fig11} as building
blocks, one can construct a large variety of knots and
links. The polynomial of the knot or link obtained by
gluing some of the balls of figures \ref{fig10} and
\ref{fig11} can be evaluated by substituting eqs. 
(\ref{ctvseis}), (\ref{ctvnueve}) and (\ref{cttuno}) in
eqs. (\ref{ctcinco}) and (\ref{ctcatorce}). In most of the
cases there exists more that one way of constructing a
given knot or link. This fact can be used to generate many
relations which, in particular, allow to determine the
matrix $a_{rl}$ uniquely. Let us see some examples. First
of all, it is a simple exercise to verify graphically that
the link $\hat L_0(\hat Q_{2m}^H, \hat Q_{2p}^H)$ is the
same as $\hat {\cal L}_{2m+2p}$. The equality of the
corresponding polynomials requires that:
\beq
\sum_{{l=0\atop 2l\in\ZZ}}^{2j}\,
a_{rl}\,a_{nl}\,=\,\delta_{rn}\,\,.
\label{cttdos}
\eeq
On the other hand, it is obvious that 
$\hat L_0(\hat Q_{2m}^H, \hat Q_{2p}^V)$ and 
$\hat L_0(\hat Q_{2p}^H, \hat Q_{2m}^H)$ are the same
link. However, the corresponding polynomials are equal only
if the $a_{rl}$ matrix is symmetric, \ie\ if:
\beq
a_{rl}\,=\,a_{lr}\,\,.
\label{ctttres}
\eeq
Taken together eqs. (\ref{cttdos}) and (\ref{ctttres}) imply
that $a_{rl}$ is a symmetric orthogonal matrix. Moreover, it
is easy to verify that 
$\hat L_0\,(\,\hat Q^{\,H}_{2m}, \hat Q^{\,V}_{0}\,)$ is
nothing but the unknot.  The requirement: 
\beq
P_j[\,\hat L_0\,(\,\hat Q^{\,H}_{2m}, \hat
Q^{\,V}_{0}\,)\,](q)
\,=\,P_j[\,{\rm unknot}\,](q)\,=\,1\,\,,
\label{cttcuatro}
\eeq
is fulfilled only if the $a_{rl}$ matrix satisfies:
\beq
\sum_{l=0\atop 2l\,\in\,\ZZ}^{2j}\,
i^{2(l-<l>)}\,
\sqrt{\Bigl[\,{4l+1\over 2}\,\Bigr]_+}\,\,\,a_{nl}\,=\,
\delta_{n,0}\,
\Bigl[\,{4j+1\over 2}\,\Bigr]_+\,\,.
\label{cttcinco}
\eeq
It is not difficult to find a solution of eqs. 
(\ref{cttdos}) , (\ref{ctttres}) and (\ref{cttcinco}) in
terms of the ${\rm osp}(1\vert 2)$ fusion matrix 
$F_{rl}^{j}$ of eq. (\ref{sseis}). Actually, one can check
that:
\beq
a_{rl}\,=\,(-1)^{2j}\,F_{rl}^{j}\,\,,
\label{cttseis}
\eeq
satisfies our requirements. Indeed, the symmetry and
orthogonality of the matrix (\ref{cttseis}) are a
consequence of similar properties of the matrix
$F_{rl}^{j}$ (eqs. (\ref{apbuno}) and (\ref{apbdos})).
Moreover, eq. (\ref{cttcinco}) is a consequence of eq. 
(\ref{apbcuatro}). Other checks of the solution 
(\ref{cttseis}) for $a_{rl}$, some of them highly
non-trivial, are presented in appendix C.

Once the duality matrix $a_{rl}$ is determined, we can
evaluate the invariant polynomials for all types of knots
and links \cite{kaul}. The result we get from our building
blocks can be simply related to the 
results found in ref. \cite{kaul} 
for the su$(2)$ polynomials. In fact, what we get for
arbitrary knots and links is exactly the same relation found
in section 4 for torus links (eq. (\ref{nsiete})). If 
$P_j[L](q)$ ($\tilde P_j[L](t))$ is the isospin $j$ 
${\rm osp}(1\vert 2)$ (su$(2)$) polynomial, in the variable
$q$ (respectively,  $t$), of the link $L$, one has:
\beq
P_j[L](q)\,=\,(-1)^{2j(\nu_L-1)}\,\tilde P_{2j}[L](-q)\,\,,
\label{cttsiete}
\eeq
where $\nu_l$ is the number of components of the link $L$. 
In order to establish eq. (\ref{cttsiete}) in the formalism
of this section, one must relate the su$(2)$ and 
${\rm osp}(1\vert 2)$ braiding eigenvalues, both for
parallel and antiparallel strands. The frame corrected 
su$(2)$ braiding eigenvalues, which we shall denote by  
$\tilde\Lambda_{l,j}^{+}(t)$ and 
$\tilde\Lambda_{l,j}^{-}(t)$, are well known:
\bear
\tilde\Lambda_{l,j}^{+}(t)\,&=&\,
(-1)^{2j-l}\,t^{2j(j+1)-{l(l+1)\over 2}}\rc
\tilde\Lambda_{l,j}^{-}(t)\,&=&\,
(-1)^{l}\,t^{{l(l+1)\over 2}}\rc
l\,&=&\,0,\cdots,2j,\,\,\,\,\,\,\,\,\,\,\,\,\,\,\,\,\,
l\in\ZZ\,\,.\label{cttocho}\\
\nonumber
\eear
It is easy to verify that, when $t$ is identified with
$-q$, the ${\rm osp}(1\vert 2)$ and su$(2)$ eigenvalues are
related as:
\bear
\Lambda_{l,j}^{+}(q)\,&=&\,(-1)^{2j}\,
\tilde\Lambda_{2l,2j}^{+}(-q)\rc
\Lambda_{l,j}^{-}(q)\,&=&\,
\tilde\Lambda_{2l,2j}^{-}(-q)\rc
l\,&=&\,0,\cdots,2j,\,\,\,\,\,\,\,\,\,\,\,\,\,\,\,\,\,
2l\in\ZZ\,\,.\label{cttnueve}\\
\nonumber
\eear
In the process of demonstrating the statement contained in 
(\ref{cttsiete}), it is also necessary to use the relation
between the su$(2)$ and ${\rm osp}(1\vert 2)$  CS
duality matrices. Let us denote the former by 
$\tilde a_{r,l}(t)$. As proved in ref. \cite{kaul} 
$\tilde a_{r,l}(t)$ is given by:
\beq
\tilde a_{r,l}(t)\,=\,\tilde F_{rl}^j(t)\,\,,
\label{ctcuarenta}
\eeq
where $\tilde F_{rl}^j(t)$ is the su$(2)$ fusion matrix of
eq. (\ref{sesenta}). Therefore, as the 
su$(2)$ and ${\rm osp}(1\vert 2)$ fusion matrices are
related as in eq. (\ref{sdos}), it follows that:
\beq
a_{rl}\,=\,(-1)^{2j}\,\tilde a_{2r,2l}(-q)\,\,.
\label{ctcuno}
\eeq
\begin{figure}
\centerline{\hskip.4in \epsffile{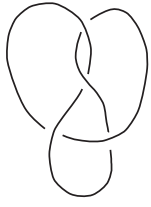}}
\caption{Representation of the figure eight knot.}
\label{fig12}
\end{figure}

Using eqs. (\ref{cttnueve}), (\ref{ctcuno}) and the
relation between the su$(2)$ and ${\rm osp}(1\vert 2)$
q-numbers (eq. (\ref{vcuatro})),  one 
can  prove eq. (\ref{ctsiete}) case by case. 
The su$(2)$ polynomials for non-trivial knots and links,
obtained from four-strand braids,
carrying arbitrary isospin representations 
have been tabulated in ref. \cite{kaul} and can be compared
with our results. In order to illustrate this
comparison, we explicitly compute the ${\rm osp}(1\vert 2)$ 
polynomial for the first simplest knot which is
not a torus knot: the figure eight knot. This knot,
which in the mathematics literature is usually denoted by
$4_{{\scriptstyle 1}}$, has been depicted in figure
\ref{fig12}. It is not difficult to verify that the knot of
figure \ref{fig12} can be obtained by gluing together the
rooms $\hat Q_2^V$ and 
$\hat Q_{-2}^H$. Indeed, the $4_{{\scriptstyle 1}}$ knot is
nothing but what we were calling 
$\hat L_0(\hat Q_2^V, \hat Q_{-2}^H)$ and, thus, we can
write:
\beq
P_j[\,4_{{\scriptstyle 1}}\,](q)\,=\,
{1\over SD_q[j]}\,
\sum_{{l=0\atop 2l\in \ZZ}}^{2j}\,
\mu_l^{(-)}(\hat Q_2^V)\,\mu_l^{(-)}(\hat Q_{-2}^H)\,\,.
\label{ctcdos}
\eeq
Taking the values of $\mu_l^{(-)}(\hat Q_2^V)$ and
$\mu_l^{(-)}(\hat Q_{-2}^H)$ from eq. (\ref{cttuno}) and
substituting them in the right-hand side of eq. 
(\ref{ctcdos}), one arrives at:
\bear
P_j[\,4_{{\scriptstyle 1}}\,](q)\,&=&\,
{1\over SD_q[j]}\,
\sum_{{l=0\atop 2l\in \ZZ}}^{2j}\,
\sum_{{r=0\atop 2r\in \ZZ}}^{2j}\,
i^{2(l-<l>)}\,i^{2(r-<r>)}\,\times\rc\rc
&&\times\sqrt{\Bigl[\,{4l+1\over 2}\,\Bigr]_+}\,
\sqrt{\Bigl[\,{4r+1\over 2}\,\Bigr]_+}\,
a_{lr}\,\Bigl (\,\Lambda_{l,j}^{(-)}(q)\,\Bigr)^{2}\,
\Bigl (\,\Lambda_{r,j}^{(-)}(q)\,\Bigr)^{-2}\,\,.\rc
\label{ctctres}
\eear
Moreover, since $SD_q[j]=(-1)^{2j}[4j+1]$ when $t=-q$ (see
eqs. (\ref{vdos}) and (\ref{vcinco})), and making use of
eqs. (\ref{vcuatro}), (\ref{cttnueve}) and (\ref{ctcuno}),
it is straightforward to verify that eq. (\ref{ctctres})
can be put in the form:
\beq
P_j[\,4_{{\scriptstyle 1}}\,](q)\,=\,
{1\over [4j+1]}\,
\sum_{{l=0\atop l\in \ZZ}}^{4j}\,
\sum_{{r=0\atop r\in \ZZ}}^{4j}\,
\sqrt{[\,4l+1\,]}\,
\sqrt{[\,4r+1\,]}\,
\tilde a_{l,r}(-q)\,\Bigl
(\,\tilde\Lambda_{l,2j}^{(-)}(-q)\,\Bigr)^{2}\, \Bigl
(\,\tilde \Lambda_{r,2j}^{(-)}(-q)\,\Bigr)^{-2}\,\,,
\label{ctccuatro}
\eeq
which, indeed, shows that 
$P_j[\,4_{{\scriptstyle 1}}\,](q)\,=\,
\tilde P_{2j}[\,4_{{\scriptstyle 1}}\,](-q)$, in agreement
with eq. (\ref{cttsiete}). 

\setcounter{equation}{0}
\section{ Summary and conclusions}

We have studied the behaviour of 
the conformal blocks under the crossing symmetry operations
in osp$(1\vert 2)$ CFT. We have concentrated our efforts in
the four-point correlators on the two-sphere. Our main tool
has been the Coulomb gas representation of the conformal
blocks which is obtained in the free field realization of
the  osp$(1\vert 2)$ current algebra. We  have found closed
expressions for the braiding and fusion matrices which, at
least for the equal isospin case studied in section 3, are
very similar to the ones corresponding to the su$(2)$ CFT.

As mentioned at the end of section 2, the $q=-t$
correspondence between osp$(1\vert 2)$ and su$(2)$  has
been previously noticed in the context of the quantum group
theory.  The fact that our fusion and braiding matrices can
also be connected to their su$(2)$ counterparts by means of
this identification (see eqs. (\ref{sdos}) and
(\ref{setenta})) is an indication of their quantum group
origin, \ie\ it can be considered as a clue that reveals
the existence of a hidden $U_q({\rm  osp}(1\vert 2))$
symmetry in the model studied. In order to confirm this
conclusion, one should verify whether or not the fusion
matrix of the osp$(1\vert 2)$ CFT is given by the $6j$
symbols of  $U_q({\rm  osp}(1\vert 2))$. It is
interesting to point out in this respect that an analytical
expression of the $6j$ symbols of  
$U_q({\rm  osp}(1\vert 2))$ has been recently reported in
ref. \cite{MM}. This expression is very similar to the one
we have found for the fusion matrix (eq. (\ref{sseis})).
However, it has been obtained with normalization
conventions different from the ones we have adopted here.
Therefore, the comparison between the  $6j$ symbols of ref. 
\cite{MM} and eq. (\ref{sseis}) cannot be done easily and,
as a consequence, more work is required in order to reach a
firm conclusion about this subject. 

Our analysis of section 4 has served to verify explicitly
the connection between the osp$(1\vert 2)$ CS theory and
the corresponding CFT. It is interesting to notice that
this connection is not exactly the same as in the su$(2)$
case. Indeed, we have found that the osp$(1\vert 2)$ Wilson
line operators are not, in general, Verlinde operators
(they can differ in a sign, see eq. (\ref{ocuatro})).  This
difference, which is crucial in order to reproduce the
fusion algebra in the space of the characters, could shed
light in the analysis of the validity of the Verlinde
theorem in other non-unitary CFTs. Notice that one can
interpret the sign appearing in the right-hand side of eq. 
(\ref{nsiete}) as coming from the relative sign between the
Verlinde and Wilson line operators in eq. (\ref{ocuatro}). 

It is also interesting to point out the consistency between
the genus one analysis of section 4 and the genus zero
formalism of section 5. Actually, both approaches are
complementary since the information obtained with the knot
operators can be used to fix the parameters of the genus
zero formalism. It is interesting to notice the highly
non-trivial checks that the duality matrix of eq.
(\ref{ctvocho}) must satisfy. The fact that we were able to
find a consistent solution for this duality matrix (eq. 
(\ref{cttseis})) is a new confirmation of the correctness
of our ansatz of eq. (\ref{sseis}) for the fusion matrix.

There are, of course, many possible extensions of our work.
The most obvious one is the analysis of the invariants
corresponding to multicoloured links, \ie\ to links with a
different osp$(1\vert 2)$ representation in each of their 
components. This analysis requires to extend our study of
the two-dimensional crossing symmetry to a more general
class of conformal blocks. It would be interesting to see if
there also exists in this case a relation with the su$(2)$
results. On the other hand, this analysis could be the
starting point in the formulation of a quantum topology
program that could lead to the discovery of new invariants
for three-manifolds in the line of ref. \cite{QT}. Within
the quantum group approach the  osp$(1\vert 2)$ invariants
for three-manifolds have been considered in ref.
\cite{Zhangdos}. In the CS framework one must compute the
vacuum expectation values of tetrahedral configurations of
Wilson lines (see the second paper of ref. \cite{witten}).
The values of this tetrahedra should be related to the
$6j$ symbols of
$U_q({\rm  osp}(1\vert 2))$ and, presumably, also to the
fusion matrix of our  osp$(1\vert 2)$ CFT.

\setcounter{equation}{0}
\section{ Acknowledgements}
P. R. would like to thank the Department of Particle
Physics of the University of Santiago, where this work was
initiated, for hospitality. We would also like to thank J. M.
F. Labastida for discussions. This
work was supported in part by DGICYT under grant PB93-0344, 
by CICYT under grant  AEN96-1673 and by the European Union
TMR grant ERBFMRXCT960012.

\newpage

\vskip 1cm                                               
{\Large{\bf APPENDIX A}}                                 
\vskip .5cm                                              
\renewcommand{\theequation}{\rm{A}.\arabic{equation}}  
\setcounter{equation}{0}  

Let us consider the integrals appearing in the vacuum
expectation values of the type (\ref{tcinco}) for
$j_1=j_2=1/2$. In the free field representation described
in the main text, these correlators are given by contour
integrals that involve the function:

\bear
\eta(\tau_1,\tau_2)&=&-\Bigl\{\,
<\,\chi(0)\,\chi(1))\,
(w(\infty))^{1+s}\,w(\tau_1)\,>\,
<\,\psi(\tau_1)\bar\psi(\tau_{2})\,>\,+\,\rc\rc
&&+<\,\chi(0)\,\chi(1))\,
(w(\infty))^{1+s}\,w(\tau_2)\,>\,
<\,\bar\psi(\tau_1)\psi(\tau_{2})\,>\,\Bigr\}\,\,.
\label{apauno}\\
\nonumber
\eear
(Compare eq. (\ref{apauno}) with eq. (\ref{tocho})). The
function (\ref{apauno}) is multiplied in the integrand of
the free-field representation of (\ref{tcinco}) by
functions of the type:

\bear
f_1(\tau_1,\tau_2)\,&=&\,\tau_1^a\,\tau_2^a\,(\tau_1-z)^b\,(\tau_2-z)^b\,
(\tau_1-1)^b\,(\tau_2-1)^b\,(\tau_1-\tau_2)^{2\rho}\rc\rc
f_2(\tau_1,\tau_2)\,&=&\,\tau_1^a\,\tau_2^a\,(\tau_1-z)^b\,(z-\tau_2)^b\,
(\tau_1-1)^b\,(1-\tau_2)^b\,(\tau_1-\tau_2)^{2\rho}
\label{apados}\\\rc
f_3(\tau_1,\tau_2)\,&=&\,\tau_1^a\,\tau_2^a\,(z-\tau_1)^b\,(z-\tau_2)^b\,
(1-\tau_1)^b\,(1-\tau_2)^b\,(\tau_1-\tau_2)^{2\rho}\,\,.\rc
\nonumber
\eear
The phases of the functions (\ref{apados}) have been chosen
in agreement with eq. (\ref{cuarenta}). Let us denote by 
${\cal I}_p(z)$ the integrals defining the blocks  
${}^s{\cal F}^{1234}_p(z)$. The  ${\cal I}_p$'s are given
by ordered contour integrals (see eq. (\ref{tnueve})).
Closely related to these functions are the integrals:

\bear
I_1(z)\,&=&\,z^{2\rho}\,(1-z)^{2\rho}\,
\int_{1}^{\infty}\,du_1\,\int_{1}^{\infty}du_2\,
f_1(u_1,u_2)\,\,\eta(u_1,u_2)\,\rc\rc
I_2(z)\,&=&\,z^{2\rho}\,(1-z)^{2\rho}\,
\int_{1}^{\infty}\,du_1\,\int_{0}^{z}dv_1\,
\,f_2(u_1, v_1)\,\eta(u_1,v_1)\label{apatres}\\\nonumber\\
I_3(z)\,&=&\,z^{2\rho}\,(1-z)^{2\rho}\,
\int_{0}^{z}\,dv_1\,\int_{0}^{z}dv_2\,
f_3(v_1,v_2)\,\eta(v_1,v_2)\,\,.\rc
\nonumber
\eear
The only difference between the  ${\cal I}_p$'s and the 
 $I_p$'s is the fact that in the former all double
integrations in a given interval are ordered, whereas, as
shown in eq. (\ref{apatres}), no such an ordering appears in
the expression that defines the  $I_p$'s. It is easy to
prove that these two types of integrals are proportional to
each other. In fact one has:
\bear
I_1(z)\,&=&2e^{i\pi(\rho-{1\over 2})}\,
c(\rho-{1\over 2})\,{\cal I}_1(z)\rc
I_2(z)\,&=&{\cal I}_2(z)\label{apacuatro}\\
I_3(z)\,&=&2e^{i\pi(\rho-{1\over 2})}\,
c(\rho-{1\over 2})\,{\cal I}_3(z)\,\,,\rc
\nonumber
\eear
where $c(x)\equiv cos \pi x$. It will be convenient in what
follows to use the functions $I_p(z)$ rather than the
ordered integrals ${\cal I}_p(z)$. The contours of
integration in eq. (\ref{apatres}) are the ones shown in
figure 2 for the particular case $n=2$. Notice that, in
order to simplify the notation, we have suppressed the
$p$-dependence of the function 
$\eta_p(\{u_i\},\{v_i\})$ in the right-hand side of eq. 
(\ref{apatres}). 

In order to obtain the matrix elements of the braiding
matrix ${\cal B}^{{1\over 2}}$, one has to relate the
integrals ${\cal I}_p(z)$ to the $s$-channel blocks
corresponding to the correlator:
\beq
{\cal F}^{1324}(z)\,\equiv\,
<\,\Phi^{{1\over 2}}_{-{1\over 2}}(0)\,
\Phi^{{1\over 2}}_{-{1\over 2}}(1)\,
\Phi^{{1\over 2}}_{{1\over 2}}(z)
\,\tilde\Phi^{{1\over 2}}_{{1\over 2}}(\infty)\, Q^{2}\,>\,.
\label{apacinco}
\eeq
\begin{figure}
\centerline{\hskip.4in \epsffile{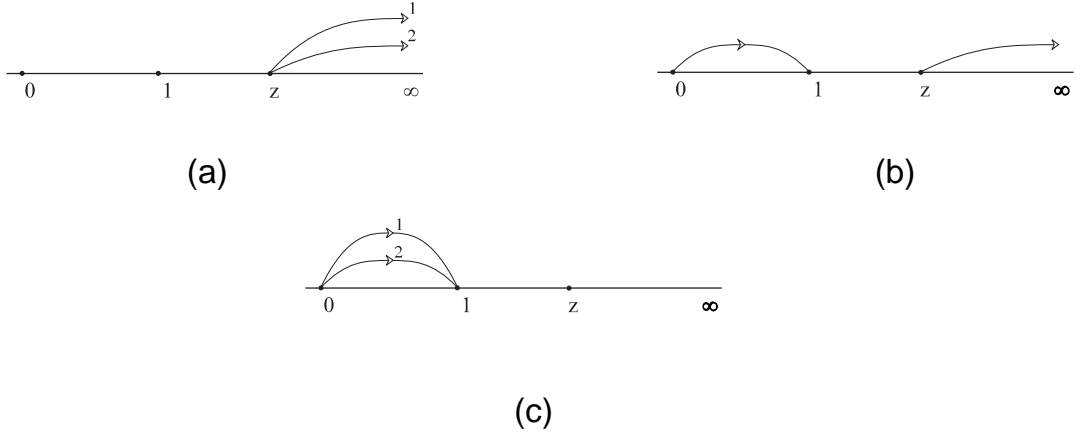}}
\caption{Contours of integration used in the definition of 
${\cal J}_p(z)$ and $J_p(z)$ for $p=1$(a), $p=2$(b) and 
$p=3$(c).}
\label{figa1}
\end{figure}
The integral appearing in the free field representation of
the block ${}^s{\cal F}^{1324}_p(z)$ will be denoted by 
${\cal J}_p(z)$. In this case, the contours of integration
differ from the ones of figure \ref{fig2} in the exchange
of the points $\tau=1$ and $\tau=z$. The contours
corresponding to the s-channel intermediate states of the
correlator (\ref{apacinco}) are shown in figure
\ref{figa1}. As in the case of figure \ref{fig2}, the
integrals along the contours of figure \ref{figa1} can be
ordered or not. The ordered ones are those appearing in the
free field blocks, \ie\ in the ${\cal J}_p$'s, 
while the integrals that are not ordered will be denoted by 
$J_p(z)$. The actual expressions of the latter are:
\bear
J_1(z)\,&=&\,z^{2\rho}\,(z-1)^{2\rho}\,
\int_{z}^{\infty}\,du_1\,\int_{z}^{\infty}du_2\,
f_1(u_1,u_2)\,\eta(u_1,u_2)\,\rc\rc
J_2(z)\,&=&\,z^{2\rho}\,(z-1)^{2\rho}\,
\int_{z}^{\infty}\,du_1\,\int_{0}^{1}dv_1\,
f_2(u_1, v_1)\,\eta(u_1,v_1)\label{apaseis}\\\rc
J_3(z)\,&=&\,z^{2\rho}\,(z-1)^{2\rho}\,
\int_{0}^{1}\,dv_1\,\int_{0}^{1}dv_2\,
f_3(v_1,v_2)\,\eta(v_1,v_2)\,\,.\rc
\nonumber
\eear
The relation between  the $J_p$ and ${\cal J}_p$ integrals
is similar to the one written in eq. (\ref{apacuatro}),
namely:
\bear
J_1(z)\,&=&2e^{i\pi(\rho-{1\over 2})}\,
c(\rho-{1\over 2})\,{\cal J}_1(z)\rc
J_2(z)\,&=&{\cal J}_2(z)\label{apasiete}\\
J_3(z)\,&=&2e^{i\pi(\rho-{1\over 2})}\,
c(\rho-{1\over 2})\,{\cal J}_3(z)\,\,.\rc
\nonumber
\eear
\begin{figure}
\centerline{\hskip-.4in \epsffile{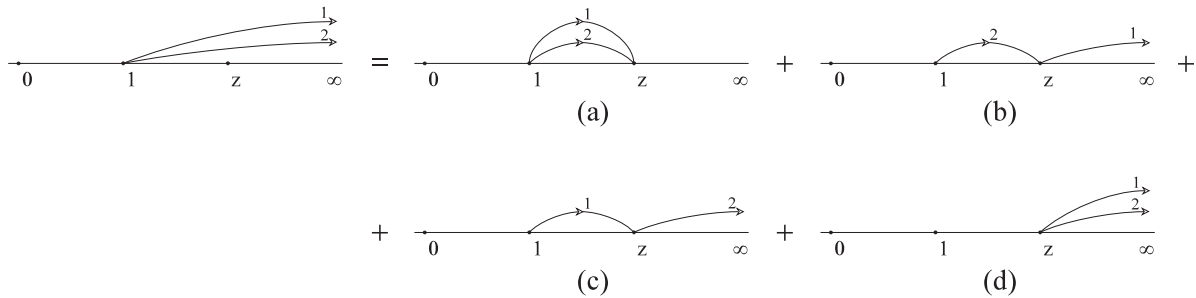}}
\caption{Decomposition of $I_1(z)$ for $z>1$ into a sum of
integrals in the intervals $(1,z)$ and $(z,\infty)$.}
\label{figa2}
\end{figure}

The phase convention used to define the integrals 
$I_p(z)$ and ${\cal I}_p(z)$ corresponds to the situation
in which $z<1$, which is the ordering of these two points
in the contours of figure \ref{fig2}. On the contrary, as
shown in figure \ref{figa1}, one should define the
integrals $J_p(z)$ and ${\cal J}_p(z)$ with a phase
convention adapted to the ordering in which $z$ is greater
than one. In order to relate the integrals ${\cal I}_p(z)$
and ${\cal J}_p(z)$ one must, first of all, analytically
continue the ${\cal I}_p(z)$'s to the region $z>1$. After
this is done, these two integrals are related linearly as
follows:
\beq
{\cal I}_p(z)\,=\,\sum_{l=1}^{3}\,c_{pl}\,{\cal J}_l(z)\,\,.
\label{apaocho}
\eeq
Taking into account the definitions of the functions 
${\cal I}_p(z)$ and ${\cal J}_p(z)$ and of the braiding
matrix (eq. (\ref{vocho})), it is clear that:
\beq
{\cal B}^{{1\over 2}}_{j_p\,j_l}\,=\,c_{pl}\,\,,
\label{apanueve}
\eeq
where the relation between the isospin and the number of
contours is given in eq. (\ref{cuno}).

As mentioned above, in order to obtain the explicit value
of the coefficients $c_{pl}$, it is more convenient to deal
with the integrals (\ref{apatres}) and (\ref{apaseis}). The
relation between these two types of integrals can be
obtained by using contour 
manipulation techniques \cite{DF}. We shall
illustrate this procedure by showing in detail how the 
$I_1(z)$ integral can be put in terms of $J_1(z)$, 
$J_2(z)$ and $J_3(z)$. First of all,  one must assume that in
$I_1(z)$, the variable $z$ is analytically continued to the
region $z>1$. The function $I_1(z)$ is defined by means of
a double integration in the interval $(1,\infty)$. As shown
in figure \ref{figa2}, each of these two integrals in the
interval $(1,\infty)$ can be represented as the sum of two
integrals performed along the $(1,z)$ and $(z,\infty)$
intervals. One gets in this way four contributions of the
type:
\beq
I_1(z)\,=\,I_1^{(a)}(z)\,+\,I_1^{(b)}(z)\,+\,I_1^{(c)}(z)\,
+\,I_1^{(d)}(z)\,\,,
\label{apadiez}
\eeq
where the superindices $a$, $b$, $c$ and $d$ refer to the
four contributions displayed in figure \ref{figa2}. In all
the integrals of the right-hand side of eq. (\ref{apadiez}),
the integrand is the same as in $I_1(z)$. It is clear by
inspection  that the integral $I_1^{(d)}(z)$ is of the same
type of $J_1(z)$ (see figure \ref{figa1}a). Taking into
account the different conventions for the relative phase of
$z$ and $1$, one can write:
\beq
I_1^{(d)}(z)\,=\,e^{2i\pi \rho}\,J_1(z)\,\,.
\label{apaonce}
\eeq
The other three remaining integrals are defined along
contours that do not correspond to any of those shown in
figure \ref{figa1}. It is in these cases where the
manipulation of contours is needed. Let us explain in
detail this technique for the integral $I_1^{(b)}$. In this
case, the variable which is not integrated over the
intervals represented in  figure \ref{figa1} (\ie\ 
$(0,1)$ and $(z,\infty)$) is the one we have called $2$.
This variable is integrated in $I_1^{(b)}$ over the interval 
$(1,z)$. By ``opening" the contour, one can convert this
integral into an integral in which the variable $2$ runs
over the intervals  $(\infty, 1)$ and $(z,\infty)$. One
must, however, be careful with the phases that are picked
up when the branch points of the integrand are crossed. In
fact, we have two different possibilities to open the
contour of the variable $2$. If we open the contour from
above, we get the following representation of $I_1^{(b)}$:
\bear
\centerline{\hskip-.4in \epsffile{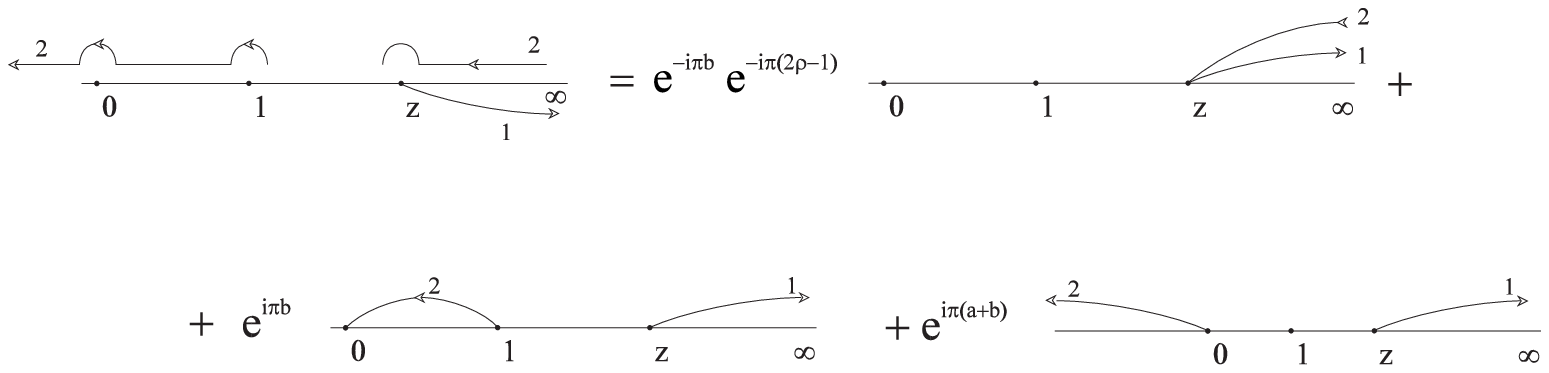}}
\rc\rc
\label{apadoce}
\eear
while, on the contrary, opening the contour of the variable 
2 from below, we obtain:
\bear
\centerline{\hskip-.4in \epsffile{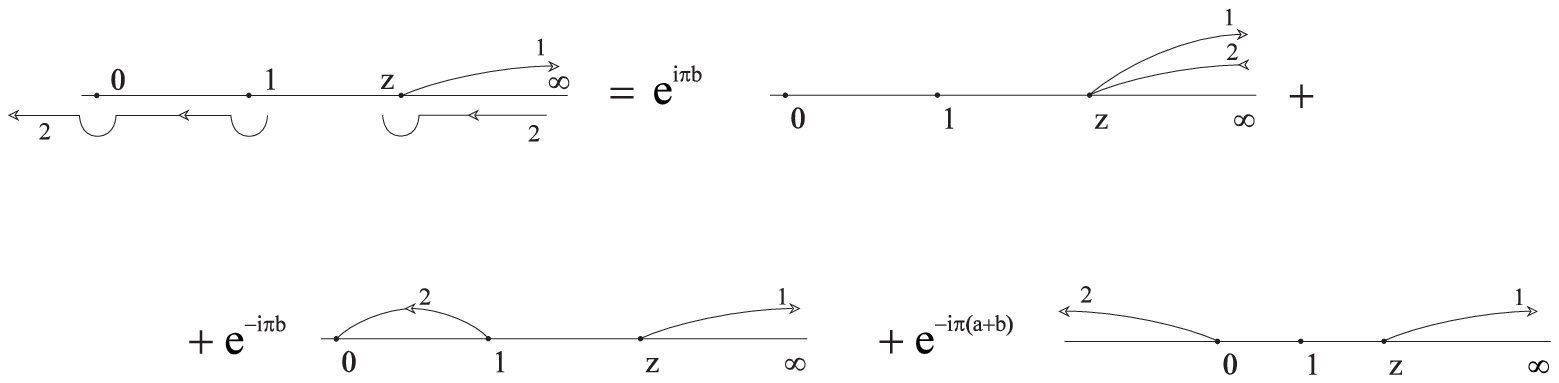}}
\rc\rc
\label{apatrece}
\eear
In eqs. (\ref{apadoce}) and (\ref{apatrece}) the left-hand
side is the same. The first two integrals of the right-hand
side of these equations correspond to some of the contours
of figure \ref{figa1},  whereas the third term does not
coincide with any of the integration paths that define the
functions $J_p(z)$. In order to get rid of this unwanted
integral, we shall  multiply eq. (\ref{apadoce}) by 
$e^{-i\pi(a+b)}$ and let us subtract it from the result of
multiplying eq.  (\ref{apatrece}) by $e^{i\pi(a+b)}$. After
these manipulations, one can obtain $I_1^{(b)}(z)$ in terms
of
$J_1(z)$ and $J_2(z)$. This result is the following:
\beq
I_1^{(b)}(z)\,=\,-e^{i\pi(b+\rho+{1\over 2})}\,\,
{s(a+2b+\rho-{1\over 2})\over
s(a+b)}\,\,J_1(z)\,-\,e^{i\pi(b+2\rho)}\,\,
{s(a)\over s(a+b)}\,\,J_2(z)\,\,,
\label{apacatorce}
\eeq
where $s(x)\equiv sin\pi x$. The other contributions
appearing in the right-hand side of eq. (\ref{apadiez}) can
be treated similarly. For the integral $I_1^{(a)}(z)$ one
gets:
\bear
I_1^{(a)}(z)&=&e^{2i\pi(b+\rho)}\,\,
{s(a+2b+2\rho-1)\,s(a+2b+\rho-{1\over 2})\over
s(a+b)\, s(a+b+\rho-{1\over 2})}\,\,J_1(z)\,+\rc\rc
&&+\,e^{i\pi(2b+3\rho-{1\over 2})}\,\,
{s(a)\,s(a+2b+2\rho-1)\over s(a+b+\rho-{1\over 2})}\,
\Biggl[\,{1\over s(a+b)}\,+\,{1\over s(a+b+2\rho-1)}\,
\Biggr]\,J_2(z)\,+\rc\rc
&&+\,e^{2i\pi(b+\rho)}\,\,
{s(a)\,s(a+\rho-{1\over 2})
\over s(a+b+\rho-{1\over 2})\,s(a+b+2\rho-1)}\,J_3(z)
\,\,,\rc\rc
\label{apaquince}
\eear
while $I_1^{(c)}(z)$ can be written as:
\beq
I_1^{(c)}(z)\,=\,-e^{i\pi(b+3\rho-{1\over 2})}\,\,
{s(a+2b+\rho-{1\over 2})\over s(a+b)}\,\,J_1(z)\,-\,
e^{i\pi(b+4\rho-1)}\,\,
{s(a)\over s(a+b)}\,\,J_2(z)\,\,.
\label{apadseis}
\eeq
Putting together the results of eqs. (\ref{apaonce}), 
(\ref{apacatorce}), (\ref{apaquince})  and
(\ref{apadseis}), and taking into account the relation with
the ordered integrals (eqs.  (\ref{apacuatro}) and
(\ref{apasiete})), one obtains the following values for the
matrix elements $c_{1l}$:
\bear
c_{11}&=&e^{2i\pi(a+2b+2\rho-{1\over 2})}\,\,
{s(b)\,s(b+\rho-{1\over 2})\over 
s(a+b+\rho-{1\over 2})\,s(a+b)}\rc\rc
c_{12}&=&e^{i\pi(a+3b+4\rho-1)}\,\,
{s(a)\,s(b)\over 
s(a+b+2\rho-1)\,s(a+b)}\label{apadsiete}\\\rc
c_{13}&=&e^{2i\pi(b+\rho)}\,\,
{s(a)\,s(a+\rho-{1\over 2})\over 
s(a+b+\rho-{1\over 2})\,s(a+b+2\rho-1)}\,\,.\rc
\nonumber
\eear 
Using these techniques, although in some cases the
calculations are much more involved, the other elements of
the braiding matrix can be obtained. The result is:
\bear
c_{21}&=&2\,e^{i\pi(a+3b+4\rho-1)}\,\,\,
{s(a+2b+\rho-{1\over 2})\,s(b+\rho-{1\over 2})\,\,
c(\rho-{1\over 2}) \over
s(a+b)\,s(a+b+\rho-{1\over 2})}\rc\rc
c_{22}&=&e^{i\pi(2b+4\rho-1)}\,\,\,
{s(a)\,s(a+2b+2\rho-1)\,-\,[s(b)]^2\over
s(a+b)\,s(a+b+2\rho-1)}\rc\rc
c_{23}&=&-2\,e^{i\pi(b-a+2\rho)}\,\,\,
{s(b+\rho-{1\over 2})\,s(a+\rho-{1\over 2})
\,\,\,c(\rho-{1\over 2}) \over
s(a+b+\rho-{1\over 2})\,s(a+b+2\rho-1)}
\label{apadocho}\\\rc
c_{31}&=&e^{2i\pi(b+\rho)}\,\,\,
{s(a+2b+2\rho-1)\,s(a+2b+\rho-{1\over 2})\over
s(a+b)\,s(a+b+\rho-{1\over 2})}\rc\rc
c_{32}&=&-\,e^{i\pi(b-a+2\rho)}\,\,\,
{s(a+2b+2\rho-1)\,s(b)\over
s(a+b)\,s(a+b+2\rho-1)}\rc\rc
c_{33}&=&e^{-2i\pi(a-{1\over 2})}\,\,\,
{s(b+\rho-{1\over 2})\,s(b)\over
s(a+b+\rho-{1\over 2})\,s(a+b+2\rho-1)}\,\,.\rc\rc
\nonumber
\eear
Substituting $a=b=-\alpha_0^2=-2\rho$ in eqs.
(\ref{apadsiete}) and (\ref{apadocho}), and taking eq.
(\ref{apanueve}) into account, one arrives at the matrix 
${\cal B}^{{1\over 2}}$ of eq. (\ref{cicinco}).

\newpage

\vskip 1cm                                               
{\Large{\bf APPENDIX B}}                                 
\vskip .5cm                                              
\renewcommand{\theequation}{\rm{B}.\arabic{equation}}  
\setcounter{equation}{0}  

In this appendix we shall collect some properties of the
duality matrix $F^j$. Some of these properties can be
obtained directly from  eq. (\ref{sseis}), while others can
be easily derived  from the identification (\ref{sdos})
between the ${\rm osp}(1\vert 2)$ and su$(2)$ fusion
matrices. 

It is evident after inspecting eq. (\ref{sseis}) that $F^j$
is a symmetric matrix, namely  one has:
\beq
F_{j_1\,j_2}^{j}\,=F_{j_2\,j_1}^{j}\,\,.
\label{apbuno}
\eeq
Moreover, $F^j$ is an orthogonal matrix, \ie\ it satisfies:
\beq
\sum_{j_3}\,F_{j_1\,j_3}^{j}\,\,
F_{j_2\,j_3}^{j}\,=\,\delta_{j_1,j_2}\,\,.
\label{apbdos}
\eeq
By means of a direct substitution in eq. (\ref{sseis}), one
can obtain the  value of the first row (or column):
\beq
F_{0\,j_1}^{j}\,=\,(-1)^{2j}\,\,
i^{\,2(j_1-<j_1>)}\,\,\,
{\sqrt{\Bigl[\,{4j_1+1\over 2}\Bigr]_+}\over
\Bigl[\,{4j+1\over 2}\Bigr]_+}\,\,.
\label{apbtres}
\eeq
Substituting the above value  in the orthogonality
condition (eq. (\ref{apbdos})), one gets:
\beq
\sum_{j_2}\,i^{\,2(j_2-<j_2>)}\,\,
\sqrt{\Bigl[\,{4j_2+1\over 2}\Bigr]_+}\,\,
F_{j_1\,j_2}^{j}\,=\,
(-1)^{2j}\,\Bigl[\,{4j+1\over 2}\Bigr]_+\,\,
\delta_{j_1,0}\,\,.
\label{apbcuatro}
\eeq
Let the Casimir of the isospin $j$ representation  be
defined as in the main text, namely 
$c_j\,=\,j\,(\,j+{1\over 2}\,)$. The fusion matrix also
satisfies:
\beq
\sum_{j_3}\,(-1)^{<j_3>}\,\,
q^{\pm 2c_{j_3}\mp 8c_j}\,\,F_{j_3\,j_1}^{j}\,
F_{j_3\,j_2}^{j}\,=\,
(-1)^{<j_1>+<j_2>}\,q^{\mp 2c_{j_1}\mp 2c_{j_2}}\,\,
F_{j_1\,j_2}^{j}(q)\,\,.
\label{apbcinco}
\eeq
Eq. (\ref{apbcinco}) can be proved from a similar
relation verified by the su$(2)$ fusion matrix. 
In particular, taking $j_2=0$ and using the value of 
$F_{0\,j_1}^{j}$,  one arrives at:
\bear
&&\sum_{j_3}\,(-1)^{<j_3>}\,\,
i^{\,2(j_3-<j_3>)}\,\,
q^{\pm 2c_{j_3}\mp 8c_j}\,\,
\sqrt{\Bigl[\,{4j_3+1\over 2}\Bigr]_+}\,
F_{j_3\,j_1}^{j}\,=\,\rc
&&=\,(-1)^{<j_1>}\,\,q^{\mp 2c_{j_1}}\,\,
i^{\,2(j_1-<j_1>)}\,\,
\sqrt{\Bigl[\,{4j_1+1\over 2}\Bigr]_+}\,\,.
\label{apbseis}\\
\nonumber
\eear

\newpage

\vskip 1cm                                               
{\Large{\bf APPENDIX C}}                                 
\vskip .5cm                                              
\renewcommand{\theequation}{\rm{C}.\arabic{equation}}  
\setcounter{equation}{0}  

In this appendix we verify that our solution
(\ref{cttseis}) for the CS  duality matrix  
satisfies several consistency relations, which constitutes
a confirmation of the correctness of our result
(\ref{cttseis}) for $a_{rl}$. The relations we are going to
check can be obtained by following the same procedure as
the one used in ref. \cite{kaul} for the su$(2)$ case. Some
of the possible checks are satisfied as a consequence of the
orthogonality and symmetry properties of the $a_{rl}$ matrix.
For example, the link $L_0(Q_{2p+1}^H, Q_{2m+1}^H)$ is the
same as the link $\hat {\cal L}_{2p+2m+2}$. It is easy to
prove that the corresponding polynomials are the same if 
the matrix $a_{rl}$ is orthogonal. The verification of other
consistency conditions requires the use of more specific
properties of the ${\rm osp}(1\vert 2)$ fusion matrix, such
as the ones listed in appendix B. So, for instance, as:
\beq
L_0(\,Q_1^H, Q_m^V\,)\,=\,{\cal L}_{m+1}\,\,,
\label{apacuno}
\eeq
one must have:
\bear
&&\sum_{l,r=0\atop 2l, 2r\,\in\,\ZZ}^{2j}\,\,
i^{2(r-<r>)}\,i^{2(l-<l>)}\,
\sqrt{\Bigl[\,{4r+1\over 2}\,\Bigr]_+}\,\,\,
\sqrt{\Bigl[\,{4l+1\over 2}\,\Bigr]_+}\,\,\,
\Lambda_{r,j}^{(-)}(q)\,\,(\,\Lambda_{l,j}^{(+)}(q)\,)^{m}\,
a_{rl}\,=\,\rc\rc
&&=\,\sum_{l=0\atop 2l\,\in\,\ZZ}^{2j}\,\,
(-1)^{2l}\,\Bigl[\,{4l+1\over 2}\,\Bigr]_+\,\,
(\,\Lambda_{l,j}^{(+)}(q)\,)^{m+1}\,\,.\label{apcdos}\\
\nonumber
\eear
Let us prove that the matrix (\ref{cttseis}) satisfies eq.
(\ref{apcdos}). If we substitute 
$\Lambda_{r,j}^{(-)}(q)\,=\,(-1)^{<r>}\,q^{2c_r}$ and 
$a_{rl}\,=\,(-1)^{2j}\,F_{rl}^j$ in the left-hand side of
eq.  (\ref{apcdos}), the sum in $r$ can be evaluated with
the help of eq. (\ref{apbseis}):
\bear
&&\sum_{r=0\atop  2r\,\in\,\ZZ}^{2j}\,\,
(-1)^{<r>}\,
i^{2(r-<r>)}\,
\sqrt{\Bigl[\,{4r+1\over 2}\,\Bigr]_+}\,\,\,
\,q^{2c_r}\,\,a_{rl}\,=\,\rc
&&=\,(-1)^{2j+<l>}\,q^{8c_j-2c_l}\,i^{2(l-<l>)}\,
\sqrt{\Bigl[\,{4l+1\over 2}\,\Bigr]_+}\,=\rc
&&=\,\Lambda_{l,j}^{(+)}(q)\,i^{2(l-<l>)}\,
\sqrt{\Bigl[\,{4l+1\over 2}\,\Bigr]_+}\,\,,
\label{apctres}\\
\nonumber
\eear
where, in the last step, we have taken into account that 
$\Lambda_{l,j}^{(+)}(q)\,=\,(-1)^{2j+<l>}\,q^{8c_j-2c_l}$.
Using this result in the left-hand side of eq.
(\ref{apcdos}), one verifies the consistency required. 

Another non-trivial check of our solution for the CS duality
matrix has its origin in the link equivalence:
\beq
L_0(\,Q_{2p+1}^H, Q_1^V\,)\,=\,\hat{\cal L}_{2p+2}\,\,,
\label{apaccuatro}
\eeq
which implies the following equation for $a_{rl}$:
\bear
&&\sum_{l,r=0\atop 2l, 2r\,\in\,\ZZ}^{2j}\,\,
i^{2(r-<r>)}\,i^{2(l-<l>)}\,
\sqrt{\Bigl[\,{4r+1\over 2}\,\Bigr]_+}\,\,\,
\sqrt{\Bigl[\,{4l+1\over 2}\,\Bigr]_+}\,\,\,
(\,\Lambda_{r,j}^{(-)}(q)\,)^{2p+1}\,
\,\Lambda_{l,j}^{(+)}(q)\,
\,\,a_{rl}\,=\,\rc\rc
&&=\,\sum_{r=0\atop 2r\,\in\,\ZZ}^{2j}\,\,
(-1)^{2r}\,\Bigl[\,{4r+1\over 2}\,\Bigr]_+\,\,
(\,\Lambda_{r,j}^{(-)}(q)\,)^{2p+2}\,\,.
\label{apccinco}\\
\nonumber
\eear
Taking  $a_{rl}\,=\,(-1)^{2j}\,F_{rl}^{j}$ and 
$\Lambda_{l,j}^{(+)}(q)\,=\,(-1)^{2j+<l>}\,
q^{8c_j-2c_l}\,\,$, and using again eq. (\ref{apbseis}), 
the sum in $l$ in the left-hand side of
eq. (\ref{apccinco}) can be performed:
\bear
\sum_{l=0\atop 2l\,\in\,\ZZ}^{2j}\,\,
(-1)^{<l>}\,i^{2(l-<l>)}\,&q^{8c_j-2c_l}\,
\sqrt{\Bigl[\,{4l+1\over 2}\,\Bigr]_+}\,\,\,
F_{rl}^{j}\,=\,(-1)^{<r>}\,q^{2c_r}\,\,
i^{2(r-<r>)}\,
\sqrt{\Bigl[\,{4r+1\over 2}\,\Bigr]_+}\,=\,\rc\rc
&=\,i^{2(r-<r>)}\,\Lambda_{r,j}^{(-)}(q)\,
\sqrt{\Bigl[\,{4r+1\over 2}\,\Bigr]_+}\,\,,
\label{apcseis}\\
\nonumber
\eear
and, as a consequence,  eq. (\ref{apccinco}) is satisfied by
the  CS duality matrix (\ref{cttseis}).


\begin{thebibliography}{99}

\bibitem{MS} G. Moore and N. Seiberg, {\sl \pl} {\bf
B212} (1988) 451; {\sl \cmp} {\bf 123} (1989) 177.

 \bibitem{Review} For a review see J. Fuchs, {\sl
``Affine Lie algebras and quantum groups"}, (Cambridge
University Press, Cambridge, 1992),   S. Ketov, {\sl
``Conformal Field Theory"}, (World Scientific,
Singapore, 1995) and C. Gomez, M. Ruiz-Altaba and G.
Sierra, {\sl ``Quantum groups in two-dimensional
physics"}, (Cambridge University Press, Cambridge, 1996). 
 

\bibitem{AGS} L. Alvarez-Gaum\'e, C. Gomez and G. Sierra, 
{\sl \pl} {\bf B220} (1989) 142; 
{\sl \np} {\bf B319} (1989) 155; 
{\sl \np} {\bf B330} (1990) 347. 

\bibitem{KR} A. N. Kirillov and N. Yu. Reshetikhin, 
`` Representation of the algebra $U_q(sl(2))$,\break
q-orthogonal polynomials and invariants of links", in 
{\sl ``Infinite dimensional Lie algebras and groups"},
ed. V. G. Kac (World Scientific, Singapore, 1989).

\bibitem{Res}  N. Yu. Reshetikhin, ``Quantized universal
enveloping algebras and invariants of links I, II", 
LOMI preprints E-4-87 and E-17-87.


  
\bibitem{witten} E. Witten, 
{\sl \cmp} {\bf 121} (1989) 351; 
{\sl Nucl. Phys.} {\bf B322} (1989) 629. 



\bibitem{ber}M. Bershadsky and H.
Ooguri, {\sl \pl} {\bf B229} (1989) 374.


\bibitem{polyakov}A. M. Polyakov and A. B.
Zamolodchikov, {\sl \mpl} {\bf A3} (1988) 1213.



\bibitem{osp} I. P. Ennes, A. V. Ramallo and J. M. Sanchez
de Santos, {\sl \pl} {\bf B389} (1996) 485;  
{\sl \np} {\bf B491 [PM]} (1997) 574.





\bibitem{Horne} J. H. Horne, {\sl \np} {\bf B334} (1990)
669.

\bibitem{KS} L. H. Kauffman and H. Saleur, 
{\sl \cmp} {\bf 141} (1991) 293; 
{\sl \ijmp} {\bf A7} (1992) 493.

\bibitem{Rozansky} L. Rozansky and  H. Saleur, 
{\sl \np} {\bf B376} (1992) 461.

\bibitem{Deguchi} 
T. Deguchi, {\sl \jpsc} {\bf 58} (1989) 3441; 
T. Deguchi and Y. Akutsu, 
{\sl J. Phys.} {\bf A}: {\sl Math. Gen.} {\bf 23} 
(1990) 1861.

\bibitem{Zhang}
R. B. Zhang, M. D. Gould and A. J. Bracken,
{\sl \cmp} {\bf 137} (1991) 13; 
R. B. Zhang, {\sl \jmp} {\bf 32} (1991) 2605; 
R. B. Zhang, {\sl \jmp} {\bf 33} (1992) 3918; 
J. R. Links, M. D. Gould and R. B. Zhang,  
{\sl \rmp} {\bf 5} (1993) 345; 
M. D. Gould, I. Tshohantjis and A. J. Bracken,
{\sl \rmp} {\bf 5} (1993) 533; 
J. R. Links and R. B. Zhang,
{\sl \jmp} {\bf 35} (1994) 1377.











\bibitem{LR} J.M.F. Labastida and A.V. Ramallo, 
{\sl Phys. Lett.} {\bf B227} (1989) 92;
{\sl Phys. Lett.} {\bf B228} (1989) 214;   
J.M.F. Labastida, P. M. Llatas and A.V. Ramallo, 
{\sl \np} {\bf B348} (1991) 651.





\bibitem{kaul} R.K.~Kaul and T.R. Govindarajan, 
{\sl \np} {\bf B380} (1992) 293; 
{\sl \np} {\bf B393} (1993) 392;   
P. Ramadevi, T. R. Govindarajan and R.K. Kaul, 
{\sl \np} {\bf B402} (1993) 548; 
R. K. Kaul, {\sl \cmp} {\bf 162} (1994) 289. 


 
 

\bibitem{DF} Vl.S.Dotsenko and V. A. Fateev, 
{\sl \np} {\bf B240} (1984) 312;
{\sl\np} {\bf B251} (1985) 691; {\sl \pl} {\bf B154}
 (1985) 291.


\bibitem{Verlinde} E. Verlinde, 
{\sl \np} {\bf B300} (1988) 360.



\bibitem{Pais}A. Pais and V.
Rittenberg, {\sl \jmp} {\bf 16} (1975) 2063;  M.
Scheunert, W. Nahn and  V.
Rittenberg, {\sl \jmp} {\bf 18} (1977) 155.




\bibitem{yudos}J. B. Fan and M. Yu, {\sl ``Modules over
affine Lie superalgebras"} , Academia Sinica preprint
AS-ITP-93-14, hep-th/9304122.

\bibitem{DV} R. Dijkgraaf and E. Verlinde, 
{\sl \np\ (Proc. Suppl. )} {\bf 5B} (1988) 87.




\bibitem{Kulis}P. Kulish and N.
Reshetikhin,  {\sl \lmp} {\bf 18} (1989) 143. 


\bibitem{Saleur} H. Saleur, 
{\sl \np} {\bf B336} (1990) 363.













\bibitem{knot} See, for example, L. H. Kauffman, 
{\sl ``Knots and Physics" } (World Scientific, Singapore,
1991).



\bibitem{poly} J.M.~Isidro, J.M.F.~Labastida and A.V.
~Ramallo, {\sl \np} {\bf B398} (1993) 187.


\bibitem{AW} Y. Akutsu and M. Wadati,
{\sl \jpsc} {\bf 56} (1987) 839; Y. Akutsu, T. Deguchi and M.
Wadati,  {\sl \jpsc} {\bf 56} (1987) 3464; 
{\sl \jpsc} {\bf 57} (1988) 757; for a review see 
M. Wadati, T. Deguchi and Y. Akutsu, 
{\sl \phyrep} {\bf 180} (1989) 247.


\bibitem{millet} W.B.R.~Lickorish and K.C.~Millet, {\sl
\topo} {\bf 26} (1987) 107.


\bibitem{MM} P. Minnaert and M. Mozrzymas, 
{\sl \jmp} {\bf 38} (1997) 2676.


\bibitem{QT} V. Turaev and O. Viro, 
{\sl \topo} {\bf 31} (1992) 865; 
N. Reshetikin and V. Turaev, 
{\sl \im} {\bf 103} (1991) 547; 
L. H. Kauffman and S. Lins, 
{\sl `` Temperley-Lieb recoupling
theory and invariants of 3-manifolds"}, Annals of Math.
Studies vol. 134, ( Princeton University Press, Princeton,
1994).

\bibitem{Zhangdos} R. B. Zhang, 
{\sl \mpl} {\bf A9} (1994) 1453.



\end{thebibliography}
\end{document}